\documentclass[aps,twocolumn,english,superscriptaddress,citeautoscript,preprintnumbers,amsmath,amssymb,floatfix,footinbib]{revtex4-1}
\usepackage[breaklinks=true,colorlinks,citecolor=blue,linkcolor=blue,urlcolor=blue]{hyperref}
\usepackage{amsmath}
\usepackage{graphicx}
\usepackage{dcolumn}
\usepackage{float}
\usepackage[utf8]{inputenc}
\DeclareUnicodeCharacter{2212}{-}
\usepackage{color}
\usepackage[super]{nth}

\begin{document}
	
\title{Electronic structure and physical properties of candidate topological material GdAgGe}
\author{D. Ram}
\affiliation{Department of Physics, Indian Institute of Technology, Kanpur 208016, India}
\author{J. Singh}
\affiliation{Department of Physics, Indian Institute of Technology Hyderabad, Kandi, Medak 502 285, Telangana, India}
\author{ M. K. Hooda}
\affiliation{Department of Physics, Indian Institute of Technology, Kanpur 208016, India}
\author{O. Pavlosiuk}
\affiliation{Institute of Low Temperature and Structure Research, Polish Academy of Sciences, ul. Okolna 2, 50-422 Wroclaw, Poland}
\author{V. Kanchana}
\email{kanchana@phy.iith.ac.in}
\affiliation{Department of Physics, Indian Institute of Technology Hyderabad, Kandi, Medak 502 285, Telangana, India}
\author{Z. Hossain}
\email{zakir@iitk.ac.in}
\affiliation{Department of Physics, Indian Institute of Technology, Kanpur 208016, India}
\affiliation{Institute of Low Temperature and Structure Research, Polish Academy of Sciences, ul. Okolna 2, 50-422 Wroclaw, Poland}
\author{D. Kaczorowski}
\email{d.kaczorowski@intibs.pl}
\affiliation{Institute of Low Temperature and Structure Research, Polish Academy of Sciences, ul. Okolna 2, 50-422 Wroclaw, Poland}

\begin{abstract}
	We grew needle-shaped single crystals of GdAgGe, which crystallizes in a noncentrosymmetric hexagonal crystal structure with space group \textit{P$\overline{6}$}2\textit{m} (189). The magnetic susceptibility data for \textit{H} $\perp$ \textit{c} reveal two pronounced antiferromagnetic transitions at \textit{T}$_{N1}$ = 20 K and \textit{T}$_{N2}$ = 14.5 K. The magnetic susceptibility anomalies are less prominent for \textit{H} $\parallel$ \textit{c}. The transition at \textit{T}$_{N1}$ is accompanied by a pronounced heat capacity anomaly confirming the bulk nature of the magnetic transition. Below \textit{T}$_{N1}$, the electrical resistivity data follows a \textit{T}$^{3/2}$ dependence. In the magnetically ordered state, GdAgGe shows positive transverse magnetoresistance, which increases with decreasing temperature and increasing field, reaching a value of $\sim$ 27\% at 9 T and 10 K. The Hall resistivity data and electronic band structure calculations suggest that both the hole and electron charge carriers contribute to the transport properties. The electronic band structure displays linear band crossings near the Fermi level. The calculations reveal that GdAgGe has a nodal line with drumhead surface states coupled with a nonzero Berry phase, making it a nontrivial nodal-line semimetal.

\end{abstract}

\maketitle
\section{Introduction}	
Rare-earth based intermetallic compounds are well known for their complex and wide range of physical properties such as quantum critical point, field induced first order to second order phase transition, non-Fermi liquid behavior, crystal electric field interaction, spin and charge ordering, valence fluctuation, heavy-fermion (Kondo) behavior, charge density wave, superconductivity, etc. \cite{B_Paramanik_2014,Paramanik_2016,YbAgGe_2011,Das_2014,YbAgGeandYbPtIn,EuAuAs,EuGa4_Al4}. Recently the interest has burgeoned in the rare-earth compounds due to the observation of nontrivial topological states. A special interest is in those materials which show interplay between topology and magnetism. In these materials, the topological states are protected  by certain crystalline or point or space group symmetry as they have either broken time-reversal symmetry (TRS) or inversion symmetry (IS) or both \cite{SrMnSb2}. It was reported that broken TRS in combination with strong spin orbit coupling (SOC) leads to a large anomalous Hall effect (AHE) in antiferromagnetic Weyl semimetal GdPtBi \cite{GdPtBi}. Interestingly, besides AHE, another type of Hall effect, i.e. topological Hall effect (THE) arises in some topological magnetic materials with a noncoplanar magnetization texture, defined by their nontrivial topology coupled with local magnetization \cite{CeAlGe2020}. It was theoretically predicted that THE can occur solely from Berry phase in absence of SOC as opposed to SOC induced AHE and it is favorable for those systems having low carrier density, and large spin splitting and exchange interactions \cite{PhysRevLett2004}. However, THE can also occur in higher carrier density ferromagnetic (FM) and antiferromagnetic (AFM) materials crystallizing in different lattice structures and is sensitive to several factors such as geometrical frustrations, spin chirality and thermal fluctuations, antisymmetric Dzyaloshinskii-Moriya interaction, correlation effects, and magnetic anisotropy \cite{Wang2019,MnSi,MnGe,Surgers2014,Vistoli2019,EuAgAs,CeAlGe2020}. Several different families of materials such as V$_x$Sb$_2$Te$_3$ \cite{Wang2019}, MnSi \cite{MnSi}, MnGe \cite{MnGe}, Mn$_5$Si$_3$ \cite{Surgers2014}, Ca$_{1-x}$Ce$_x$MnO$_3$ \cite{Vistoli2019}, EuAgAs \cite{EuAgAs}, EuO \cite{EuO},  and Pr$_2$Ir$_2$O$_7$ \cite{Pr2Ir2O7} were reported to display THE.

The rare-earth based equiatomic ternary germanide compounds can be very exciting from the perspective of topological, magnetic and transport properties due to their tunable magnetic behavior, which could be manipulated by number of 4\textit{f} electrons and external magnetic fields \cite{CeAlGe2020,GdAgGe, RAgGe_1998,RAgGe2004,TmAgGe}. Recently, type–I and type-II Weyl semimetal states were theoretically predicted in some of the rare-earth based aluminium germanides, where the type of topological state can be tuned by choice of rare-earth element by breaking IS or TRS \cite{RAlGe}. The theoretical predictions made in ref. \cite{RAlGe} were later confirmed by the experimental results in PrAlGe \cite{PrAlGe2020,PrAlGe2020_2}, and CeAlGe \cite{CeAlGe2018}. Interestingly, both tetragonal compounds were found to show anomalous transport associated with nontrivial Berry phase for magnetic fields along the crystallographic \textit{c}-axis \cite{PrAlGe,PrAlGe2020}. PrAlGe exhibits large AHE with conductivity value $\sim$ 680 $\Omega^{-1}$cm$^{-1}$ \cite{PrAlGe} and CeAlGe shows THE in field range 0.4-1.5 T at low temperatures \cite{CeAlGe2020}. PrAlGe was also reported to show enhanced AHE due to large Berry curvature generated by Weyl nodes, when small magnetic fields polarize Pr local moments along the \textit{c}-axis \cite{PrAlGe2020_2}. YbAgGe is another interesting compound, which exhibits field induced quantum criticality, non-Fermi liquid behavior and anisotropic Hall effect \cite{YbAgGe2004,YbAgGe2005,YbAgGe2013}. All these interesting observations motivated us to further explore the equiatomic ternary rare-earth silver germanide systems in context of their magnetic, thermodynamic, transport and topological properties.

Here, we describe the results of our study of GdAgGe that crystallizes in hexagonal crystal structure with space group \textit{P$\overline{6}$}2\textit{m} (189) and orders antiferromagnetically below the N\'{e}el temperature, \textit{T}$_{N}$ $\approx$ 15.6 K as determined for polycrystalline samples \cite{RAgGe_1998}. We grew the GdAgGe single crystals and  investigated their physical properties by means of magnetic susceptibility, heat capacity, electrical resistivity, Hall effect, and magnetoresistance measurements as well as electronic band structure calculations.

\section{Experimental details and methods}
Single crystals of GdAgGe were grown using Pb flux. High purity elements Gd (Alfa Aesar, 99.9\%), Ag (Alfa Aesar, 99.99\%), Ge (Alfa Aesar, 99.999\%), and Pb (Alfa Aesar, 99.99\%) were weighed in a 1:1:1:10 molar ratio and put into alumina crucible. Then, the crucible was sealed inside a quartz tube under argon atmosphere. Next, the whole assembly was put into a furnace and subjected to heat treatment at 1100 $^\circ$C for 10 hours. Subsequently, the furnace was slowly cooled down to 700 $^\circ$C at a rate 2 $^\circ$C/h and at this temperature flux was removed by centrifugation. The process resulted in needle-like shiny single crystals with typical dimensions of \mbox{3 $\times$ 0.3 $\times$ 0.4 mm$^3$}.
 
The crystal structure and chemical composition of the grown crystals were checked by X-ray diffraction (XRD) on a PANalytical X'Pert PRO diffractometer with Cu-K$_{\alpha1}$ radiation, and energy-dispersive X-ray spectroscopy (EDS) performed in a JEOL JSM-6010LA scanning electron microscope. Transport measurements were carried out in a Quantum Design Physical Property Measurement System (PPMS) using standard four-probe method. Heat capacity measurements were performed using conventional relaxation method in the same PPMS platform. Magnetization measurements were carried out employing a Quantum Design Magnetic Property Measurement System (MPMS).

The first-principles calculations were performed based on the density functional theory (DFT) \cite{Hohenberg, Kohn} with the projector augmented wave (PAW) \cite{PAW} method as implemented in Vienna \textit{ab initio} simulation package ({\footnotesize VASP}) \cite{Kresse1, Kresse2}. To account for exchange-correlation effects, the generalised gradient approximation (GGA) with the Perdew-Burke-Ernzerhof (PBE) \cite {PBE} parametrization was used. The significant correlation effects of Gd-\textit{f} states were handled by using a Hubbard \textit{U} parameter (GGA+\textit{U}) of 6 eV \cite{Hubbard1, Hubbard2}. All calculations were done with a plane wave energy cutoff of 600 eV, and the energy convergence criterion was chosen to be 10$^{−6}$ eV. The geometry optimization calculations were performed with 2 $\times$ 2 $\times$ 2 supercell using dense \textit{k}-mesh as per the Monkhorst-Pack technique \cite{Monkhorst}. For surface state calculations, the {\footnotesize WANNIER90} package \cite{WANNIER90} was employed to create a tight-binding Hamiltonian based on maximally localised Wannier functions. The iterative Green's function approach, which is implemented in the {\footnotesize WANNIERTOOLS} package, was used to study topological features of the compound based on the tight-binding model \cite{WANNIERTOOLS1, WANNIERTOOLS2}.

\section{Results and discussion}
\begin{figure}
	\includegraphics[width=15.4cm, keepaspectratio]{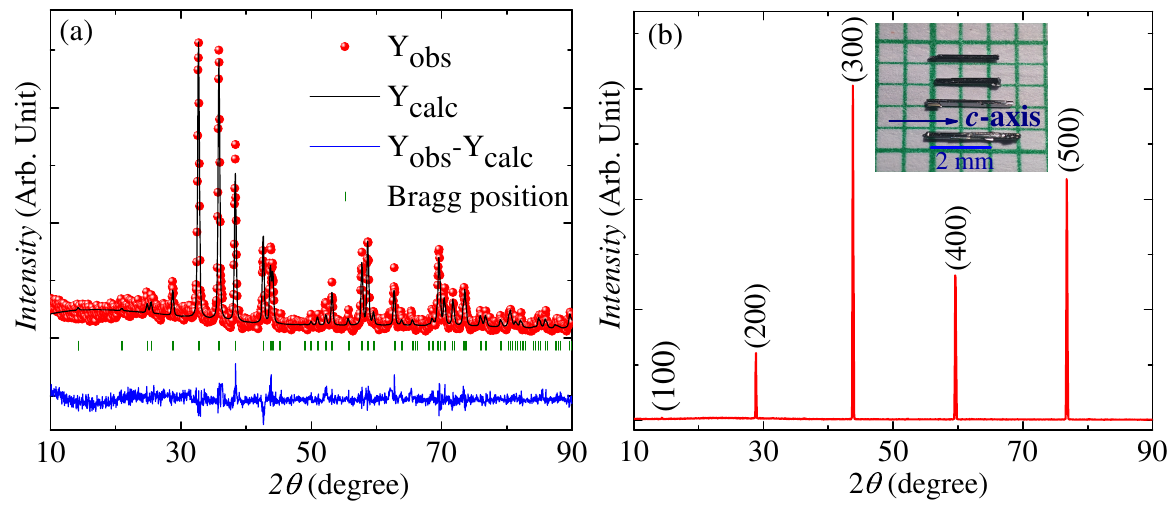}
	\caption{\label{Fig1}(a) Powder X-ray diffraction pattern of crushed single crystals of GdAgGe recorded at room temperature. (b) XRD pattern of GdAgGe single crystal. Inset shows the photograph of needle-shaped single crystals grown along the $c$-axis.}
\end{figure} 

\subsection{Crystal structure}

\begin{figure*}
	\includegraphics[width=17cm, keepaspectratio]{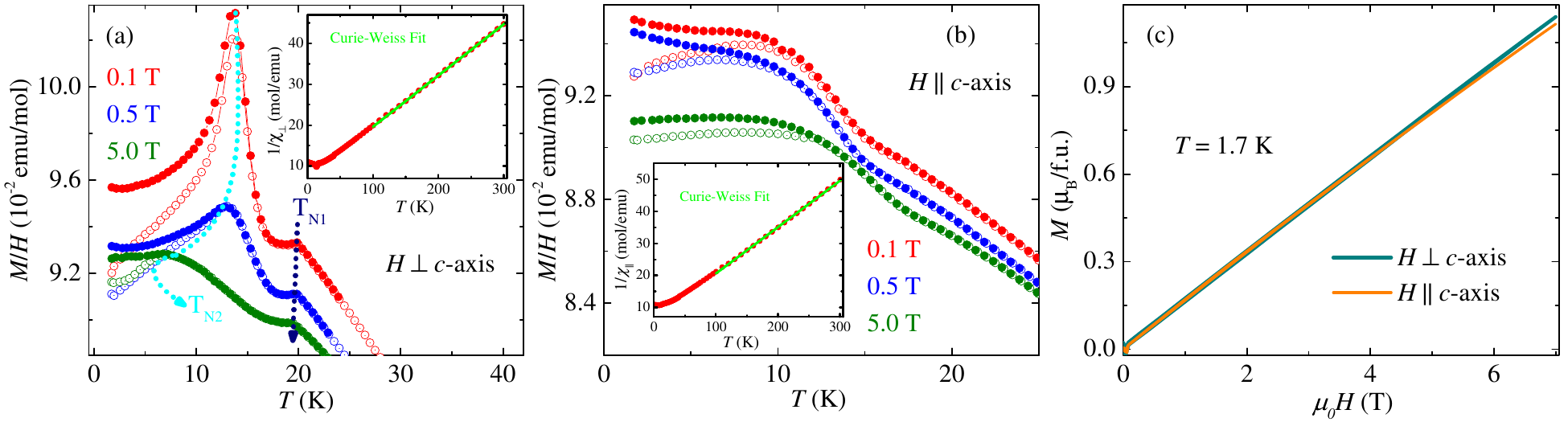}
	\caption{\label{Fig2}The magnetization data of GdAgGe single crystals recorded for fields parallel and perpendicular to the crystallographic \textit{c}-axis. Magnetization, \textit{M}/\textit{H} as function of \textit{T} measured in various fields of \textit{H} = 0.1 T, 0.5 T and 5 T in the ZFC (open circles) and FC (filled circles) modes: (a) for \textit{H} $\perp$ \textit{c} and (b) for \textit{H} $\parallel$ \textit{c}. The insets of (a) and (b) show the temperature dependence of the inverse magnetic susceptibility (1/$\chi$) for \textit{H} $\perp$ \textit{c} and \textit{H} $\parallel$ \textit{c} in a field of \textit{H} = 0.1 T, respectively. The green line is fit to Curie-Weiss law. (c) The isothermal magnetization of GdAgGe for \textit{H} $\perp$ \textit{c} and \textit{H} $\parallel$ \textit{c} at \textit{T} = 1.7 K.}
\end{figure*}
Fig. \ref{Fig1}(a) shows the powder XRD pattern of crushed single crystals of GdAgGe recorded at room temperature. It reveals the single phase growth of GdAgGe crystals. Further, we carried out Rietveld refinement of the powder XRD data using the {\footnotesize FULLPROF} software. The refinement yielded the hexagonal structure with space group \textit{P$\overline{6}$}2\textit{m} (189) and the lattice parameters \textit{a} = \textit{b} = 7.170 \AA, and \textit{c} = 4.241 \AA, close to those reported in the literature \cite{GdAgGe}. The single crystal XRD pattern measured on a flat surface of needle-shaped crystal is shown in Fig. \ref{Fig1}(b). The presence of ($l$00) peaks in the diffraction pattern indicates that a needle axis of the crystal coincides with the crystallographic \textit{c}-axis as shown in the inset of Fig. \ref{Fig1}(b). The chemical composition of needle-shaped crystals was found to be very close to the ideal equiatomic stoichiometry.

\subsection{Magnetic properties}
Figs. \ref{Fig2}(a) and \ref{Fig2}(b) show the dc magnetization (\textit{M}) as a function of temperature (\textit{T}) for various external fields (\textit{H}) applied perpendicular and parallel to the crystallographic $c$-axis. The data was taken in zero-field cooled (ZFC) and field cooled (FC) modes and it is plotted as \textit{M}(\textit{T})/\textit{H} curves. For \textit{H} $\perp$ \textit{c}, \textit{M}/\textit{H} curve shows two very sharp peaks at \textit{T}$_{N1}$ = 20 K and \textit{T}$_{N2}$  = 14.5 K, clearly indicating the AFM ordering in the compound. It is worthy to mention here that for polycrystalline sample, only one AFM transition was observed at $\sim$ 15.6 K \cite{RAgGe_1998}. As is evident from Fig. \ref{Fig2}(a), an increase in field from 0.1 to 5 T shifts the peaks towards lower temperatures, which is consistent with the usual behavior exhibited by an antiferromagnet \cite{EuZn2Ge2,RNi2Ge2}. Further, with the change in field direction from \textit{H} $\perp$ \textit{c} to \textit{H} $\parallel$ \textit{c}, the observed sharp peaks become quite broad and less pronounced. The magnetization values observed for \textit{H} $\perp$ \textit{c} and \textit{H} $\parallel$ \textit{c} do not differ significantly, suggesting small magnetic anisotropy. For both configurations, ZFC and FC curves show bifurcation below \textit{T}$_{N2}$. This feature hints at the possibility that the order-to-order magnetic phase transition in GdAgGe involves small canting of the AFM moments.

The temperature dependencies of the inverse magnetic susceptibility ($\chi^{-1}$ = \textit{M}/\textit{H}) measured in magnetic field $\mu_0$\textit{H} = 0.1 T with \textit{H} $\perp$ \textit{c} and \textit{H} $\parallel$ \textit{c} are shown in insets of Figs. \ref{Fig2}(a) and \ref{Fig2}(b), respectively. Above 100 K, the data were fitted to the Curie-Weiss equation $\chi(T) = C/(T - \Theta_P)$, where $C$ and $\Theta_P$ are the Curie constant and the paramagnetic Curie temperature, respectively. The fit yields the values $\Theta_P$ $\sim$ -56.4 K and -42.0 K for \textit{H} $\perp$ \textit{c} and \textit{H} $\parallel$ \textit{c}, respectively. The negative value of $\Theta_P$ conforms to the AFM order in the compound. The frustration parameter \textit{f} (=$\Theta_P$/\textit{T}$_{N}$) estimated for our crystal is $\sim$ 2.82 (3.89) for \textit{T}$_{N1}$ (\textit{T}$_{N2}$), which lies in moderate range \cite{Annual1994}. The \textit{f} value for GdAgGe crystal is much smaller than reported for strongly frustrated germanide YbAgGe ($\sim$26) \cite{JPS}, but larger than for polycrystalline GdAgGe ($\sim$1.77) and ErAgGe ($\sim$0.30) compounds \cite{ErAgGe}. The effective magnetic moment, $\mu_{eff}$ estimated from the Curie constant for \textit{H} $\perp$ \textit{c} (\textit{H} $\parallel$ \textit{c}) is 7.99 (7.42) $\mu_B$, which compares well with the theoretical value of $\sim$ 7.94 $\mu_B$ for Gd$^{3+}$ ion.
  
Fig. \ref{Fig2}(c) shows the isothermal magnetization, \textit{M}(\textit{H}) of GdAgGe measured at 1.7 K for \textit{H} $\perp$ \textit{c} and \mbox{\textit{H} $\parallel$ \textit{c}} configurations. \textit{M} increases linearly with \textit{H} up to 7 T for both field orientations. This kind of behavior is expected for AFM systems with metamagnetic transition occurring in stronger fields and reported before for other Gd based ternary antiferromagnets such as GdPtBi, GdAuGe and GdAuIn \cite{GdPtBi,GdAgGe}.

\subsection{Heat capacity and entropy}
\begin{figure}
	\includegraphics[width=8.0cm, keepaspectratio]{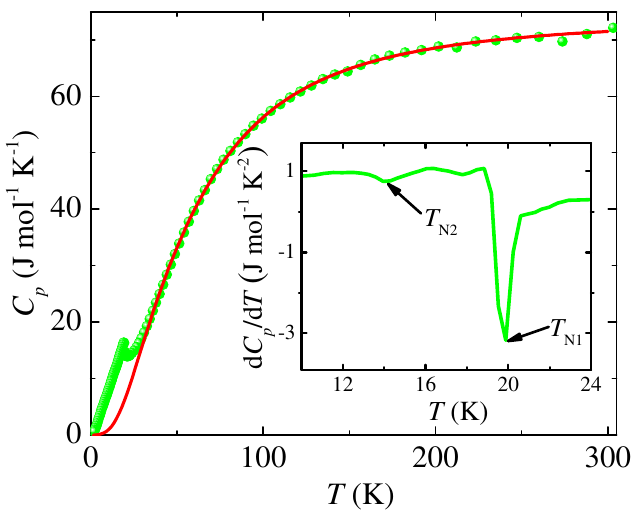}
	\caption{\label{Fig3}Temperature dependence of the heat capacity of GdAgGe at zero field. The solid red line represents the fit to the experimental data using Einstein–Debye model. The inset shows the temperature derivative of the heat capacity.}
\end{figure}

The temperature dependence of the heat capacity ($C_p$) at constant pressure is shown in Fig. \ref{Fig3}. A clear lambda-shaped anomaly near 20 K is observed, which confirms the AFM ordering at \textit{T}$_{N1}$ revealed in our magnetization measurements. The other AFM transition, clearly visible at \textit{T}$_{N2}$ = 14.5 K in the magnetic susceptibility data, does not manifest itself in $C_p$, however its weak signature is visible in the temperature derivative of the heat capacity (see the inset to Fig. \ref{Fig3}). With application of magnetic field, the anomaly at \textit{T}$_{N1}$ is slightly shifted towards lower temperatures with increasing field (not shown here), as observed in the magnetic susceptibility data. In the paramagnetic state, $C_p$ increases continuously until it gets saturated at a value of $\sim$ 72.11 J mol$^{-1}$ K$^{-1}$ at the room temperature. This value is close to the limit set by Dulong-Petit law ($C_p$ = 3\textit{n}R = 74.84 J mol$^{-1}$ K$^{-1}$, where \textit{n} is the number of atoms in the formula units, and R is the universal gas constant). The temperature variation of $C_p$ above \textit{T}$_{N1}$  can be well described by a sum of electronic ($\gamma$\textit{T}), Debye ($C_D$), and Einstein ($C_E$) contributions with the expression
\begin{equation}
C_p(\textit{T}) = \gamma \textit{T}+pC_{D}(\textit{T})+(1-p)C_{E}(\textit{T})
\label{Eq1}
\end{equation} 
where $p$ is the weight factor and $\gamma$ is the Sommerfeld coefficient. $C_D$(\textit{T}) and $C_E$(\textit{T}) are defined as
\begin{equation}
C_{D}(\textit{T})=9\textit{n}R\left( \frac{\textit{T}}{\Theta_D}\right)^3\int_{0}^{\Theta_D/\textit{T}}\frac{x^4e^x}{(e^x-1)^2}dx
\end{equation}
and
\begin{equation}
C_{E}(\textit{T})=3\textit{n}R\left( \frac{\Theta_E}{\textit{T}}\right)^2\frac{e^{\Theta_E/\textit{T}}}{(e^{\Theta_E/\textit{T}}-1)^2}
\end{equation}
where $\Theta_D$ and $\Theta_E$ are the Debye and Einstein temperatures, respectively. The obtained values of the fitting parameters are $\gamma$ = 0.47 mJ mol$^{-1}$ K$^{-2}$, $\Theta_D$ = 164 K, $\Theta_E$ = 288 K, and $p$ = 0.66. The large value of $\Theta_E$ suggests the contribution of high frequency optical modes to the heat capacity of GdAgGe.

By subtracting the phonon contribution to $C_p$(\textit{T}), determined from Eq. (\ref{Eq1}), the magnetic part of the heat capacity $C_{m}$ was derived and used to estimate the magnetic entropy according to the formula  $S_{m} = \int \frac{C_{m}}{\textit{T}}d\textit{T}$. The magnitude of $S_m$ is $\sim$ 11.5 J mol$^{-1}$ K$^{-1}$ at \textit{T}$_{N1}$ = 20 K and saturates above 30 K around 14.5 \mbox{J mol$^{-1}$ K$^{-1}$}, which is about 84 \% of the value R\textit{ln}8 expected for S = 7/2.

\subsection{Magnetotransport}
\begin{figure}
	\includegraphics[width=8.0cm, keepaspectratio]{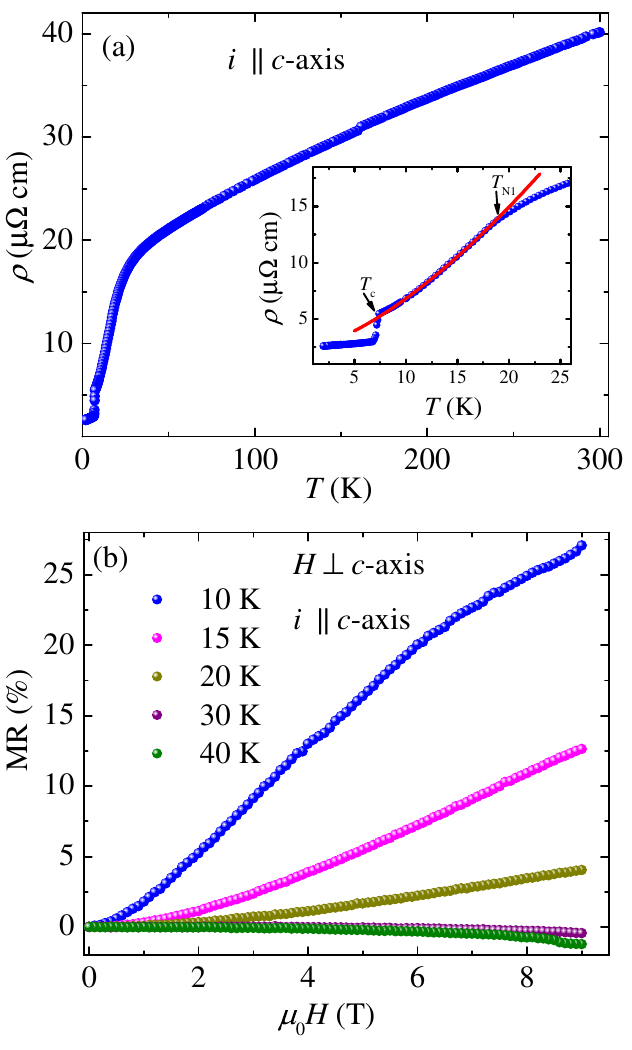}
	\caption{\label{Fig4}(a) Temperature dependence of the electrical resistivity of GdAgGe single crystals in zero magnetic field. The inset shows the expanded view of the low temperature resistivity data  with power-law fit (shown by red line) in the AFM ordering range. (b) Transverse magnetoresistance (\textit{H} $\perp$ \textit{i}) of GdAgGe single crystal measured at several temperatures.}
\end{figure}
Fig. \ref{Fig4}(a) shows the temperature dependence of the electrical resistivity $\rho$ of GdAgGe single crystal measured along the \textit{c}-axis in zero magnetic field. The resistivity decreases monotonically with decreasing \textit{T} in a metallic manner and its magnitude is similar to those reported for GdPdSn and GdPdGe \cite{GdPdX}. Below 30 K, the resistivity decreases more rapidly. The AFM transition at \textit{T}$_{N1}$ manifests itself as a knee in $\rho$(\textit{T}) and a clear minimum in the temperature derivative of this curve (not shown here). This feature can be associated with suppression of spin-disorder scattering in the magnetically ordered state. Remarkably, $\rho$(\textit{T}) does not exhibit any clear anomaly at \textit{T}$_{N2}$ = 14.5 K,  as displayed in the inset to Fig. \ref{Fig4}(a). The measured resistivity was found to drop suddenly near 7 K. This feature is likely extrinsic to GdAgGe and originates due to Pb flux residues trapped on the random sites of crystal surface \cite{Pb}, which become superconducting at this temperature. The extrinsic character of the superconductivity is confirmed by the absence of any corresponding anomaly in the heat capacity data. To determine the dominant scattering mechanism in the AFM range (8 K $\leq$ \textit{T} $\leq$ 17 K), the low temperature resistivity data were fitted with the expression $\rho$(\textit{T}) = $\rho_{0}$ + \textit{A}\textit{T}$^\textit{m}$, where $\rho_{0}$ and \textit{A} represent the residual resistivity and the scattering coefficient, respectively and \textit{m} is the exponent representing the scattering mechanism. Usually, in the AFM regime, where conduction electrons are scattered by AFM magnons, the exponent, \textit{m} = 3 is expected \cite{rossiter_1987}. In the present case, we found \textit{m} = 3/2, predicted by Moriya et al. to occur due to AFM spin fluctuations \cite{TohruMoriya1995}. Similar kind of $\rho$(\textit{T}) behavior at low temperatures was reported in the literature for RPdSi (R = Gd, Tb, Dy) compounds \cite{RPdSi}.

The isothermal transverse magnetoresistance (\textit{H} $\perp$ \textit{i}) of GdAgGe is shown in Fig. \ref{Fig4}(b), where MR is defined as MR = [$\rho$(\textit{H}) $-$ $\rho$(0)]/$\rho$(0). In the ordered state, MR is positive and increases with increasing field, reflecting the generic feature of an antiferromagnet. The maximum MR value is $\sim$ 27\% at 9 T and 10 K. At the boundary of AFM to paramagnetic phase transition i.e. at 20 K, MR magnitude sharply decreases to $\sim$ 4\% at 9 T, and becomes nearly zero in weak magnetic fields. With further increase in temperature (\textit{T} $>$ \textit{T}$_{N}$), MR is very small in small fields and becomes negative in strong fields, which can be attributed to field-induced alignment of the Gd magnetic moments in the paramagnetic phase.

\begin{figure}
	\includegraphics[width=8.0cm, keepaspectratio]{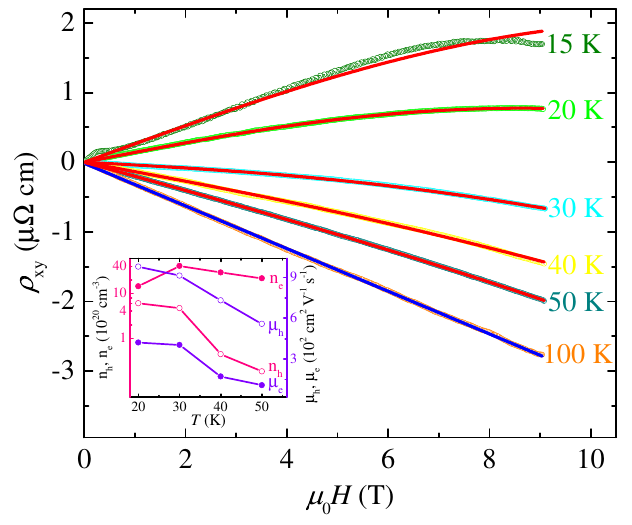}
	\caption{\label{Hall}Magnetic field dependence of Hall resistivity for GdAgGe at various temperatures. The red and blue lines represent the two-(Eq. \ref{Eq4}) and single-band model fitting, respectively. The inset shows variation of carrier concentrations and mobilities as a function of temperature.}
\end{figure}
Fig. \ref{Hall} shows the field dependence of Hall resistivity ($\rho_{xy}$) in the temperature range 15-100 K. It is clearly observed that at low temperatures (\textit{T} $<$ 100 K), $\rho_{xy}$ shows the nonlinear field dependence, indicating the multi-band character of electronic transport. However, the sign of $\rho_{xy}$ is positive below 30 K, which suggests that the holes are the dominating charge carriers. At somewhat higher temperatures (30 $\leq$ \textit{T} $<$ 100 K), the nonlinear behavior of $\rho_{xy}$ persists with the electrons dominating charge transport. For \textit{T} = 100 K, the $\rho_{xy}$ becomes linear with complete dominance of electrons. The slope of linear fit provides the Hall coefficient $R_{H}$, which is used to estimate electron carrier concentration (\textit{n}), and the Hall mobility ($\mu$) by using the relations \textit{n} = 1/(e$R_{H}$) and $\mu$ = $R_{H}$/$\rho_{xx}$(\textit{H} = 0). The estimated value of \textit{n} and $\mu$ is 2.04 $\times$ 10$^{21}$ cm$^{-3}$ and 118  cm$^{2}$ V$^{-1}$ s$^{-1}$, respectively. To further extract information about the charge carriers and their mobilities at low temperatures (\textit{T} $<$ 100 K), we use the semi-classical two-band model, where the Hall resistivity is given by the expression  
\begin{equation}
\rho_{xy}=\frac{H}{e}\frac{(n_h\mu_h^2-n_e\mu_e^2)+(n_h-n_e)\mu_e^2\mu_h^2H^2}{{(n_e\mu_e+n_h\mu_{h\ })}^2+{(n_h-n_e)}^2\mu_e^2\mu_h^2H^2}
\label{Eq4}
\end{equation}
 The two-band model fit in the temperature range 15-50 K is shown in Fig. \ref{Hall}. It is clearly evident that the two-band model nicely fits the Hall data above 20 K. However, at 15 K, the quality of fit is rather poor due to the magnetic ordering effects on the $\rho_{xy}$. From the fitting, we estimate the carrier density of $\sim$ 10$^{19}$-10$^{21}$ cm$^{-3}$ in the temperature range 15-50 K. The details of different charge carrier densities and their mobilities are shown in inset of Fig. \ref{Hall}. The estimated carrier densities in GdAgGe are significantly higher than that of typical Dirac/Weyl semimetals (\textit{n} $\sim$ 10$^{17}$-10$^{19}$ cm$^{-3}$), but comparable to that reported in several nodal-line semimetals \cite{ZrSiSe&ZrSiTe,ZrSiS,YbCdGe,Ta3SiTe6}. We do not observe the signature of THE down to 15 K. This could be due to the fact that the Dirac node is not precisely located at the Fermi level, and hence the contribution of Dirac fermions is probably masked by the dominating conventional charge carriers.

\subsection{Electronic structure}
\begin{figure}[t]
\centering
\includegraphics[width=85mm,height=38mm]{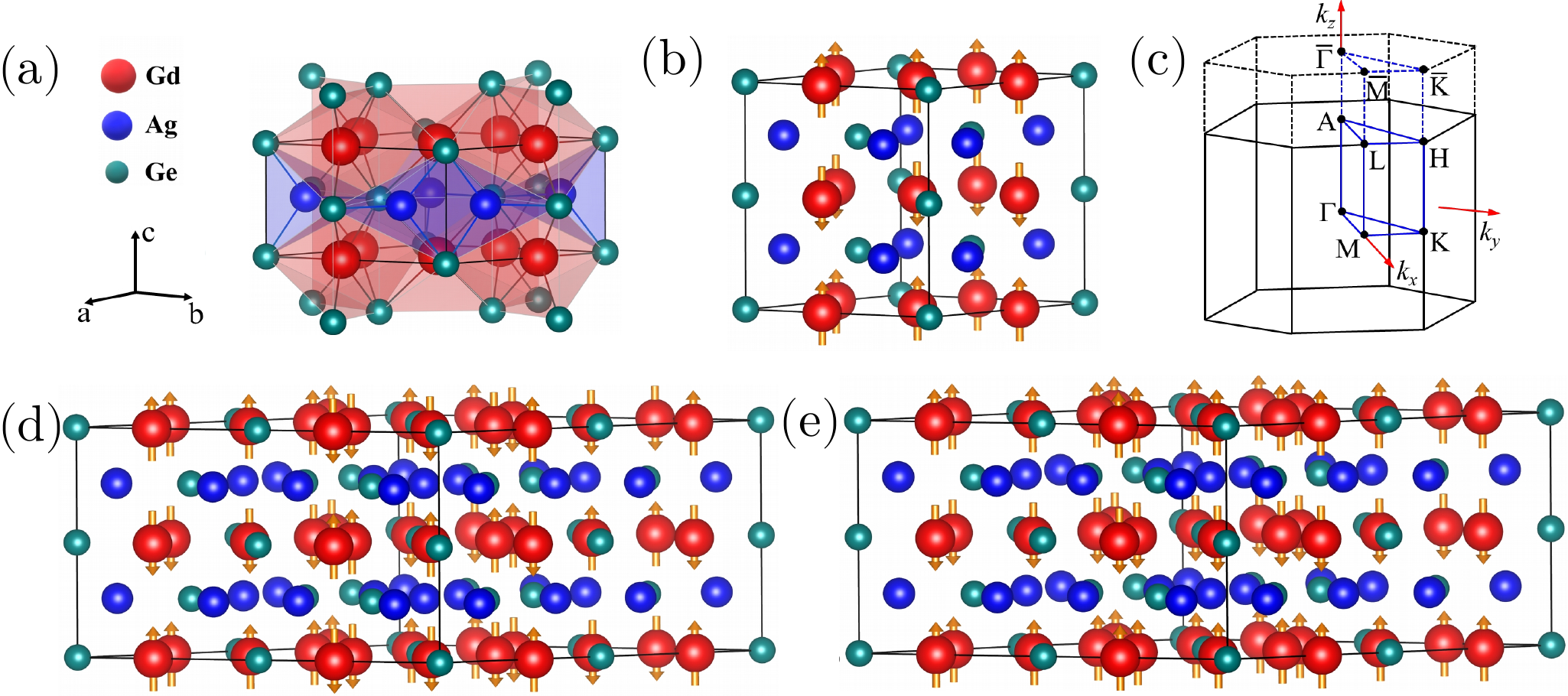}
\caption{(a) The crystal structure of GdAgGe, where Gd and Ag atoms are situated at the pyramidal and tetrahedral sites, respectively. (b) AFM configuration for 1 $\times$ 1 $\times$ 2 supercell of GdAgGe with [001] directed magnetic moments. (c) The irreducible Brillouin zone of the bulk along with the (001) projected surface. (d) AFM1 configuration for 2 $\times$ 2 $\times$ 2 supercell with in- and out of plane AFM interactions. (e) AFM2 configuration for 2 $\times$ 2 $\times$ 2 supercell with FM interactions in \textit{ab}-pane and AFM interactions along \textit{c}-axis.}
\label{structure}
\end{figure}

\begin{figure}[t]
\centering
\includegraphics[width=86mm,height=63mm]{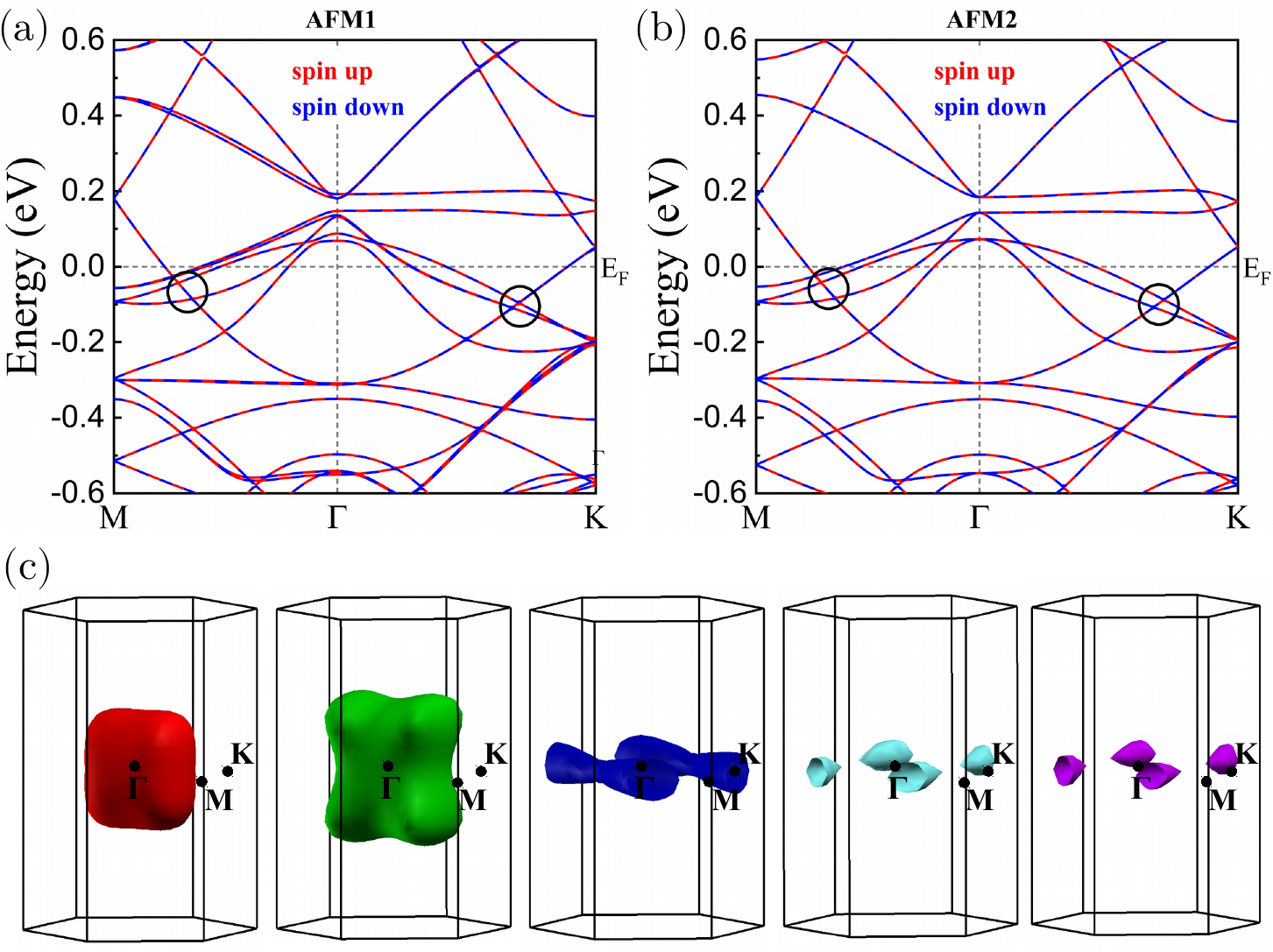}
\caption{Electronic band structure along M-$\Gamma$-K path for (a) AFM1 and (b) AFM2 configuration. (c) The Fermi surfaces in 3D-Brillouin zone without spin-orbit coupling with \mbox{2 $\times$ 2 $\times$ 2} supercell for AFM1 configuration.}
\label{AFM_bands}
\end{figure}

GdAgGe crystallizes in a hexagonal system with noncentrosymmetric space group \textit{P$\overline{6}$}2\textit{m} (189), which is structurally identical to ZrNiAl, a ternary ordered form of Fe$_2$P as illustrated in Fig. \ref{structure}(a). The space group has a threefold rotational axis (\textit{C$_{3z}$}) and a horizontal mirror plane \textit{m$_{001}$}, but lacks inversion symmetry. Our DFT calculations describe the crystal structure with the lattice parameters \textit{a} = \textit{b} = 7.24 \AA, \textit{c} = 4.26 \AA, in good agreement to those determined experimentally. Gd and Ag atoms are positioned at the pyramidal (3\textit{g}) and tetrahedral (3\textit{f}) sites, respectively. From Fig. \ref{structure}(a), it can be seen that Gd + Ge and Ag + Ge atoms form the layers that are separated along the \textit{c}-axis. According to the experimental results, GdAgGe is an AFM system. Thus, we used the 2 $ \times$ 2 $\times$ 2 supercell to compute the ground state energy for non-magnetic (NM), FM and AFM configurations to gain a better understanding of the magnetic characteristics. The possible AFM configurations are shown in Figs. \ref{structure}(d) and \ref{structure}(e). Here, AFM1 configuration contains the AFM coupling of magnetic moments in- and out of plane, whereas AFM2 (A-type AFM) configuration exhibit FM coupling in the \textit{ab}-plane and AFM interactions along the \textit{c}-axis. The calculated energy differences among the different magnetic orderings are tabulated in Table \ref{table}. As evident from Table \ref{table}, AFM1 has lowest ground state energy among the calculated structures and it is consistent with our experimental data. It should be noted that the energy difference between AFM1 and AFM2 is very small and computed electronic band structures as well as the topological features for both cases show no significant differences at the Fermi level [see Figs. \ref{AFM_bands}(a) and \ref{AFM_bands}(b)]. To further check the spin orientations of moments in AFM1 structure (which explain our experimental data), we have calculated the ground state energies for [001], [010], [100], [011], [101], [110] and [111] configurations. The calculated energy differences are given in Table \ref{SOC_table}, which shows minimum ground state energy for the [001] configuration.

\begin{table}[htb]
\caption{Calculated energies of different non-magnetic and magnetic configurations (in meV) with the reference energy considered to be 0 meV} 
\vskip .1cm
\begin{tabular}{c c c c c}
\hline\hline 	 \\[0.01ex]
Configuration ~~~~~~& NM~~~~ & FM~~~~ & AFM1~~~~& AFM2 \\ [1.5ex] 
\hline \\[0.01ex]
Energy (meV) ~~~~~~& 680~~~~&35.10~~~~ & 0 ~~~~ & 1.92 \\ [1.5ex] 
\hline
\end{tabular}
\label{table}
\end{table}

\begin{table}[htb]
\caption{Calculated energies of different spin configurations in AFM1 state with the reference energy considered to be 0 $\mu$eV} 
\begin{tabular}{c c c c c c c c}
\hline\hline\\[0.01ex]
Configuration ~& [001] & [010] & [100] & [011] & [101] & [110] & [111] \\ [1.5ex] 
\hline \\[0.01ex]
Energy ($\mu$eV)~ & 0.0 & 26.09 & 25.13 & 13.40 & 12.85 & 21.96 & 14.93 \\ [1.5ex] 
\hline
\end{tabular}
\label{SOC_table}
\end{table}

\begin{figure}[t]
\centering
\includegraphics[width=88mm,height=72mm]{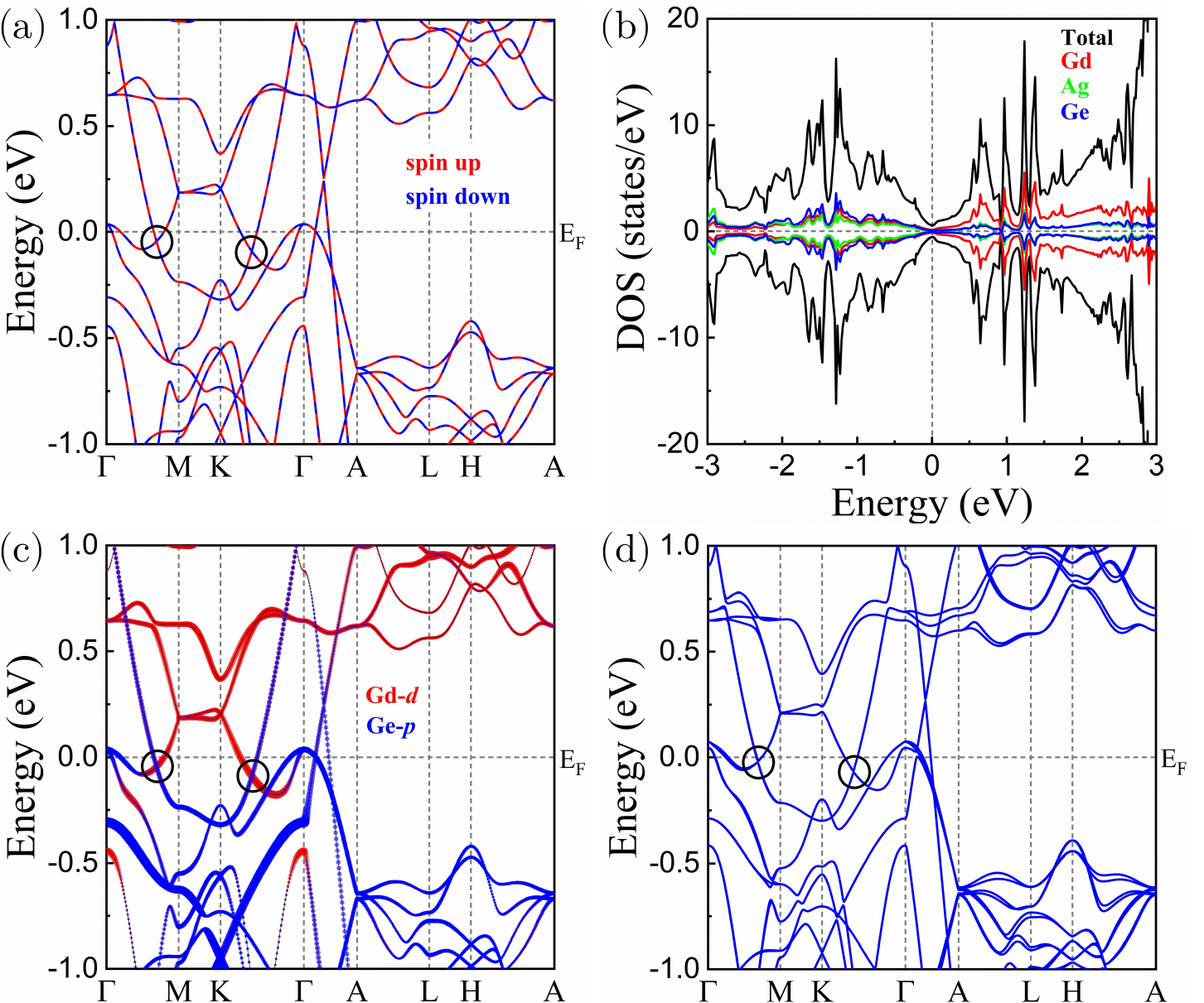}
\caption{(a) Electronic band structure along $\Gamma$-M-K-$\Gamma$-A-L-H-A path without SOC. (b) Total and projected density of states of GdAgGe. (c) The character bands Gd-\textit{d} and Ge-\textit{p} without SOC. (d) Electronic band structure with SOC along [001] direction.}
\label{bands}
\end{figure}

\begin{figure}[t]
\centering
\includegraphics[width=86mm,height=73mm]{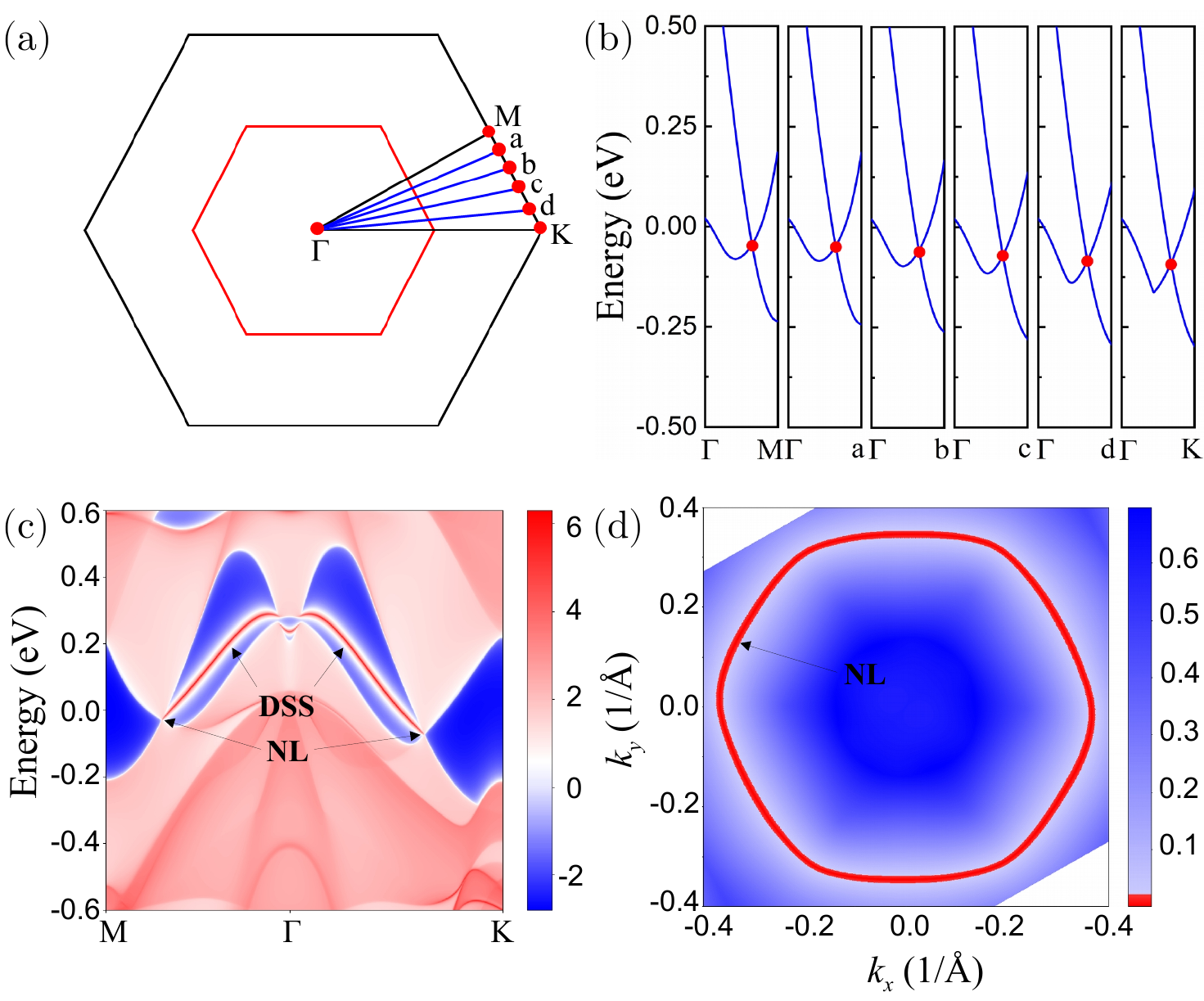}
\caption{(a) The illustration of the nodal line, where a, b, c and d are equally spaced points between M and K. (b) Electronic band structure along the \textit{k} paths as indicated in (a). (c) (001) projected surface states together with (d) energy gap plane, which show the nodal line.}
\label{nodal}
\end{figure}
As explained above, the energy difference between AFM1 and AFM2 is very minimal, and the calculated electronic band structures for both configurations look alike as shown in Figs. \ref{AFM_bands}(a) and \ref{AFM_bands}(b). Moreover, we have calculated the Fermi surfaces for AFM1 configuration, which are depicted in Fig. \ref{AFM_bands}(c). We can see that the electron and hole pockets are present at the Fermi level, which is in good agreement with the experimentally observed Hall resistivity. The investigated compound's electronic band structure displays the band crossings in \textit{k$_z$} = 0 plane and it gives us a hint towards the topological features. To further explore the topological features, we have performed the electronic bands as well as the surface analysis for lower cell (1 $\times$ 1 $\times$ 2) of AFM configuration [as shown in Fig. \ref{structure}(b)] and such type of calculations for bigger cell is beyond the scope of our work.

In the next step, we have proceeded with \mbox{1 $\times$ 1 $\times$ 2} supercell of AFM configuration. Fig. \ref{bands}(a) shows the band structure of GdAgGe with the spin-up (in red color) and the spin-down (in blue color) channel. The examined compound's electronic properties demonstrate the semi-metallic nature. Fig. \ref{bands}(b) displays the total density of states (DOS) along with atomic projected density of states. The DOS values in both the conduction and valence band region around the Fermi level are almost same, which reflects the semi-metallic nature of the compound. Interestingly, the conduction and valence bands cross each other in a linear manner along the \textit{k$_z$} = 0 plane (marked in Fig. \ref{bands}(a)). These Dirac-like crossings hint toward the presence of nodal line, which is protected by mirror symmetry located in the \textit{xy}-plane. To check for possible nontrivial nature of bands, we calculated the orbital decomposed band structure (Fig. \ref{bands}(c)), which infers that Gd-\textit{d} and Ge-\textit{p} states are main contributors to the bands forming nodal line. Close inspections revealed the Gd-\textit{d} and Ge-\textit{p} band inversion together with opposite mirror eigenvalues +1 and -1, respectively, at the $\Gamma$ point. To analyse these band crossings, we carefully examined the band structure along equally spaced paths between M and K, which is shown in Fig. \ref{nodal}(a). It can be seen that the band crossings appear along $\Gamma$-M/a/b/c/d/K paths, which infers that a $\Gamma$-centered nodal line should occur in the \textit{k$_z$} = 0 plane. To demonstrate the topological features in GdAgGe, we calculated the surface states in the (001) direction using the {\footnotesize WANNIER} package. Fig. \ref{nodal}(c) illustrates the nodal line dispersions from bulk bands and emergent drumhead surface states in the (001) projected surface state. In addition, the energy gap plane and the Fermi surface corresponding to the bands crossings also confirm the occurrence of nodal line in GdAgGe, which is illustrated in Fig. \ref{nodal}(d). The topological invariant, i.e., the Berry phase (a \textit{Z}$_2$-type invariant) along a closed path encircling the nodal line, usually protects the nodal line structure of a topological nodal-line semimetal \cite{Berry}. To confirm this point we have calculated the Berry phase and found a nonzero quantized Berry phase, which confirms GdAgGe to be a nontrivial nodal-line semimetal.

The presence of band inversion without SOC and nodal line in \textit{k$_z$} = 0 plane further lead us to include the SOC effects here. The electronic band structure with SOC is shown in Fig. \ref{bands}(d), which reflects that the band crossings are nearly unaffected. However, the inclusion of SOC opens a negligible band gap of 5.7 meV and 5.9 meV along the $\Gamma$-M and K-$\Gamma$ path, respectively. These gap values are quite smaller than those found in other reported nodal-line semimetals \cite{YbCdGe, YbCdSn, CaTX}.

\section{Conclusions}

We studied the physical properties of GdAgGe single crystals by measuring the magnetic susceptibility, heat capacity, magnetoresistance, Hall effect, and by computing the electronic band structure. The magnetic susceptibility measurements in GdAgGe showed the two successive AFM transitions at \textit{T}$_{N1}$ = 20 K and \textit{T}$_{N2}$ = 14.5 K, which are most prominent for fields perpendicular to the crystallographic \textit{c}-axis. In the heat capacity data, a clear lambda-shaped anomaly confirms the bulk character of AFM transition observed at \textit{T}$_{N1}$ = 20 K. The electrical resistivity shows a sharp drop  below \textit{T}$_{N1}$ and \textit{T}$^{3/2}$ dependence in the AFM state. The transverse magnetoresistance is positive for \textit{T} $\leq$ \textit{T}$_{N1}$, and becomes negative in the paramagnetic region. It increases with increasing field and reaches $\sim$ 27\% at 9 T and 10 K. The Hall resistivity data show the multi-band character of the compound, where hole carriers are dominating at low temperatures and electron carriers at high temperatures. The electronic band structure calculations reveal the presence of a nodal line with drumhead surface states in \textit{k$_z$} = 0 plane, which is protected by the reflection symmetry.

\section{Acknowledgment}
We acknowledge IIT Kanpur and the Department of Science and Technology, India, [Order No.
DST/NM/TUE/QM-06/2019 (G)] for financial support. VK acknowledges IIT Hyderabad for its computational facility. JS was supported through CSIR scholarship. ZH acknowledges financial support from the Polish National Agency for Academic Exchange under Ulam fellowship. DK acknowledges financial support by the National Science Centre (Poland) under research grant 2021/41/B/ST3/01141. 

\bibliography{Reference_GdAgGe}

\begin{thebibliography}{60}%
\makeatletter
\providecommand \@ifxundefined [1]{%
 \@ifx{#1\undefined}
}%
\providecommand \@ifnum [1]{%
 \ifnum #1\expandafter \@firstoftwo
 \else \expandafter \@secondoftwo
 \fi
}%
\providecommand \@ifx [1]{%
 \ifx #1\expandafter \@firstoftwo
 \else \expandafter \@secondoftwo
 \fi
}%
\providecommand \natexlab [1]{#1}%
\providecommand \enquote  [1]{``#1''}%
\providecommand \bibnamefont  [1]{#1}%
\providecommand \bibfnamefont [1]{#1}%
\providecommand \citenamefont [1]{#1}%
\providecommand \href@noop [0]{\@secondoftwo}%
\providecommand \href [0]{\begingroup \@sanitize@url \@href}%
\providecommand \@href[1]{\@@startlink{#1}\@@href}%
\providecommand \@@href[1]{\endgroup#1\@@endlink}%
\providecommand \@sanitize@url [0]{\catcode `\\12\catcode `\$12\catcode
  `\&12\catcode `\#12\catcode `\^12\catcode `\_12\catcode `\%12\relax}%
\providecommand \@@startlink[1]{}%
\providecommand \@@endlink[0]{}%
\providecommand \url  [0]{\begingroup\@sanitize@url \@url }%
\providecommand \@url [1]{\endgroup\@href {#1}{\urlprefix }}%
\providecommand \urlprefix  [0]{URL }%
\providecommand \Eprint [0]{\href }%
\providecommand \doibase [0]{http://dx.doi.org/}%
\providecommand \selectlanguage [0]{\@gobble}%
\providecommand \bibinfo  [0]{\@secondoftwo}%
\providecommand \bibfield  [0]{\@secondoftwo}%
\providecommand \translation [1]{[#1]}%
\providecommand \BibitemOpen [0]{}%
\providecommand \bibitemStop [0]{}%
\providecommand \bibitemNoStop [0]{.\EOS\space}%
\providecommand \EOS [0]{\spacefactor3000\relax}%
\providecommand \BibitemShut  [1]{\csname bibitem#1\endcsname}%
\let\auto@bib@innerbib\@empty
\bibitem [{\citenamefont {Paramanik}\ \emph {et~al.}(2014)\citenamefont
  {Paramanik}, \citenamefont {Paulose}, \citenamefont {Ramakrishnan},
  \citenamefont {Nigam}, \citenamefont {Geibel},\ and\ \citenamefont
  {Hossain}}]{B_Paramanik_2014}%
  \BibitemOpen
  \bibfield  {author} {\bibinfo {author} {\bibfnamefont {U.~B.}\ \bibnamefont
  {Paramanik}}, \bibinfo {author} {\bibfnamefont {P.~L.}\ \bibnamefont
  {Paulose}}, \bibinfo {author} {\bibfnamefont {S.}~\bibnamefont
  {Ramakrishnan}}, \bibinfo {author} {\bibfnamefont {A.~K.}\ \bibnamefont
  {Nigam}}, \bibinfo {author} {\bibfnamefont {C.}~\bibnamefont {Geibel}}, \
  and\ \bibinfo {author} {\bibfnamefont {Z.}~\bibnamefont {Hossain}},\ }\href
  {\doibase 10.1088/0953-2048/27/7/075012} {\bibfield  {journal} {\bibinfo
  {journal} {Superconductor Science and Technology}\ }\textbf {\bibinfo
  {volume} {27}},\ \bibinfo {pages} {075012} (\bibinfo {year}
  {2014})}\BibitemShut {NoStop}%
\bibitem [{\citenamefont {Paramanik}\ \emph {et~al.}(2016)\citenamefont
  {Paramanik}, \citenamefont {Bar}, \citenamefont {Das}, \citenamefont
  {Caroca-Canales}, \citenamefont {Prasad}, \citenamefont {Geibel},\ and\
  \citenamefont {Hossain}}]{Paramanik_2016}%
  \BibitemOpen
  \bibfield  {author} {\bibinfo {author} {\bibfnamefont {U.~B.}\ \bibnamefont
  {Paramanik}}, \bibinfo {author} {\bibfnamefont {A.}~\bibnamefont {Bar}},
  \bibinfo {author} {\bibfnamefont {D.}~\bibnamefont {Das}}, \bibinfo {author}
  {\bibfnamefont {N.}~\bibnamefont {Caroca-Canales}}, \bibinfo {author}
  {\bibfnamefont {R.}~\bibnamefont {Prasad}}, \bibinfo {author} {\bibfnamefont
  {C.}~\bibnamefont {Geibel}}, \ and\ \bibinfo {author} {\bibfnamefont
  {Z.}~\bibnamefont {Hossain}},\ }\href {\doibase
  10.1088/0953-8984/28/16/166001} {\bibfield  {journal} {\bibinfo  {journal}
  {Journal of Physics: Condensed Matter}\ }\textbf {\bibinfo {volume} {28}},\
  \bibinfo {pages} {166001} (\bibinfo {year} {2016})}\BibitemShut {NoStop}%
\bibitem [{\citenamefont {Schmiedeshoff}\ \emph {et~al.}(2011)\citenamefont
  {Schmiedeshoff}, \citenamefont {Mun}, \citenamefont {Lounsbury},
  \citenamefont {Tracy}, \citenamefont {Palm}, \citenamefont {Hannahs},
  \citenamefont {Park}, \citenamefont {Murphy}, \citenamefont {Bud'ko},\ and\
  \citenamefont {Canfield}}]{YbAgGe_2011}%
  \BibitemOpen
  \bibfield  {author} {\bibinfo {author} {\bibfnamefont {G.~M.}\ \bibnamefont
  {Schmiedeshoff}}, \bibinfo {author} {\bibfnamefont {E.~D.}\ \bibnamefont
  {Mun}}, \bibinfo {author} {\bibfnamefont {A.~W.}\ \bibnamefont {Lounsbury}},
  \bibinfo {author} {\bibfnamefont {S.~J.}\ \bibnamefont {Tracy}}, \bibinfo
  {author} {\bibfnamefont {E.~C.}\ \bibnamefont {Palm}}, \bibinfo {author}
  {\bibfnamefont {S.~T.}\ \bibnamefont {Hannahs}}, \bibinfo {author}
  {\bibfnamefont {J.-H.}\ \bibnamefont {Park}}, \bibinfo {author}
  {\bibfnamefont {T.~P.}\ \bibnamefont {Murphy}}, \bibinfo {author}
  {\bibfnamefont {S.~L.}\ \bibnamefont {Bud'ko}}, \ and\ \bibinfo {author}
  {\bibfnamefont {P.~C.}\ \bibnamefont {Canfield}},\ }\href {\doibase
  10.1103/PhysRevB.83.180408} {\bibfield  {journal} {\bibinfo  {journal} {Phys.
  Rev. B}\ }\textbf {\bibinfo {volume} {83}},\ \bibinfo {pages} {180408}
  (\bibinfo {year} {2011})}\BibitemShut {NoStop}%
\bibitem [{\citenamefont {Das}\ \emph {et~al.}(2014)\citenamefont {Das},
  \citenamefont {Gruner}, \citenamefont {Pfau}, \citenamefont {Paramanik},
  \citenamefont {Burkhardt}, \citenamefont {Geibel},\ and\ \citenamefont
  {Hossain}}]{Das_2014}%
  \BibitemOpen
  \bibfield  {author} {\bibinfo {author} {\bibfnamefont {D.}~\bibnamefont
  {Das}}, \bibinfo {author} {\bibfnamefont {T.}~\bibnamefont {Gruner}},
  \bibinfo {author} {\bibfnamefont {H.}~\bibnamefont {Pfau}}, \bibinfo {author}
  {\bibfnamefont {U.~B.}\ \bibnamefont {Paramanik}}, \bibinfo {author}
  {\bibfnamefont {U.}~\bibnamefont {Burkhardt}}, \bibinfo {author}
  {\bibfnamefont {C.}~\bibnamefont {Geibel}}, \ and\ \bibinfo {author}
  {\bibfnamefont {Z.}~\bibnamefont {Hossain}},\ }\href {\doibase
  10.1088/0953-8984/26/10/106001} {\bibfield  {journal} {\bibinfo  {journal}
  {Journal of Physics: Condensed Matter}\ }\textbf {\bibinfo {volume} {26}},\
  \bibinfo {pages} {106001} (\bibinfo {year} {2014})}\BibitemShut {NoStop}%
\bibitem [{\citenamefont {Bonville}\ \emph {et~al.}(2007)\citenamefont
  {Bonville}, \citenamefont {Rams}, \citenamefont {Kr{\'o}las}, \citenamefont
  {Sanchez}, \citenamefont {Canfield}, \citenamefont {Trovarelli},\ and\
  \citenamefont {Geibel}}]{YbAgGeandYbPtIn}%
  \BibitemOpen
  \bibfield  {author} {\bibinfo {author} {\bibfnamefont {P.}~\bibnamefont
  {Bonville}}, \bibinfo {author} {\bibfnamefont {M.}~\bibnamefont {Rams}},
  \bibinfo {author} {\bibfnamefont {K.}~\bibnamefont {Kr{\'o}las}}, \bibinfo
  {author} {\bibfnamefont {J.-P.}\ \bibnamefont {Sanchez}}, \bibinfo {author}
  {\bibfnamefont {P.~C.}\ \bibnamefont {Canfield}}, \bibinfo {author}
  {\bibfnamefont {O.}~\bibnamefont {Trovarelli}}, \ and\ \bibinfo {author}
  {\bibfnamefont {C.}~\bibnamefont {Geibel}},\ }\href {\doibase
  10.1140/epjb/e2007-00042-6} {\bibfield  {journal} {\bibinfo  {journal} {The
  European Physical Journal B}\ }\textbf {\bibinfo {volume} {55}},\ \bibinfo
  {pages} {77} (\bibinfo {year} {2007})}\BibitemShut {NoStop}%
\bibitem [{\citenamefont {Malick}\ \emph {et~al.}(2022)\citenamefont {Malick},
  \citenamefont {Singh}, \citenamefont {Laha}, \citenamefont {Kanchana},
  \citenamefont {Hossain},\ and\ \citenamefont {Kaczorowski}}]{EuAuAs}%
  \BibitemOpen
  \bibfield  {author} {\bibinfo {author} {\bibfnamefont {S.}~\bibnamefont
  {Malick}}, \bibinfo {author} {\bibfnamefont {J.}~\bibnamefont {Singh}},
  \bibinfo {author} {\bibfnamefont {A.}~\bibnamefont {Laha}}, \bibinfo {author}
  {\bibfnamefont {V.}~\bibnamefont {Kanchana}}, \bibinfo {author}
  {\bibfnamefont {Z.}~\bibnamefont {Hossain}}, \ and\ \bibinfo {author}
  {\bibfnamefont {D.}~\bibnamefont {Kaczorowski}},\ }\href {\doibase
  10.1103/PhysRevB.105.045103} {\bibfield  {journal} {\bibinfo  {journal}
  {Phys. Rev. B}\ }\textbf {\bibinfo {volume} {105}},\ \bibinfo {pages}
  {045103} (\bibinfo {year} {2022})}\BibitemShut {NoStop}%
\bibitem [{\citenamefont {Nakamura}\ \emph {et~al.}(2015)\citenamefont
  {Nakamura}, \citenamefont {Uejo}, \citenamefont {Honda}, \citenamefont
  {Takeuchi}, \citenamefont {Harima}, \citenamefont {Yamamoto}, \citenamefont
  {Haga}, \citenamefont {Matsubayashi}, \citenamefont {Uwatoko}, \citenamefont
  {Hedo}, \citenamefont {Nakama},\ and\ \citenamefont {Ōnuki}}]{EuGa4_Al4}%
  \BibitemOpen
  \bibfield  {author} {\bibinfo {author} {\bibfnamefont {A.}~\bibnamefont
  {Nakamura}}, \bibinfo {author} {\bibfnamefont {T.}~\bibnamefont {Uejo}},
  \bibinfo {author} {\bibfnamefont {F.}~\bibnamefont {Honda}}, \bibinfo
  {author} {\bibfnamefont {T.}~\bibnamefont {Takeuchi}}, \bibinfo {author}
  {\bibfnamefont {H.}~\bibnamefont {Harima}}, \bibinfo {author} {\bibfnamefont
  {E.}~\bibnamefont {Yamamoto}}, \bibinfo {author} {\bibfnamefont
  {Y.}~\bibnamefont {Haga}}, \bibinfo {author} {\bibfnamefont {K.}~\bibnamefont
  {Matsubayashi}}, \bibinfo {author} {\bibfnamefont {Y.}~\bibnamefont
  {Uwatoko}}, \bibinfo {author} {\bibfnamefont {M.}~\bibnamefont {Hedo}},
  \bibinfo {author} {\bibfnamefont {T.}~\bibnamefont {Nakama}}, \ and\ \bibinfo
  {author} {\bibfnamefont {Y.}~\bibnamefont {Ōnuki}},\ }\href {\doibase
  10.7566/JPSJ.84.124711} {\bibfield  {journal} {\bibinfo  {journal} {Journal
  of the Physical Society of Japan}\ }\textbf {\bibinfo {volume} {84}},\
  \bibinfo {pages} {124711} (\bibinfo {year} {2015})}\BibitemShut {NoStop}%
\bibitem [{\citenamefont {Liu}\ \emph {et~al.}(2017)\citenamefont {Liu},
  \citenamefont {Hu}, \citenamefont {Zhang}, \citenamefont {Graf},
  \citenamefont {Cao}, \citenamefont {Radmanesh}, \citenamefont {Adams},
  \citenamefont {Zhu}, \citenamefont {Cheng}, \citenamefont {Liu},
  \citenamefont {Phelan}, \citenamefont {Wei}, \citenamefont {Jaime},
  \citenamefont {Balakirev}, \citenamefont {Tennant}, \citenamefont {DiTusa},
  \citenamefont {Chiorescu}, \citenamefont {Spinu},\ and\ \citenamefont
  {Mao}}]{SrMnSb2}%
  \BibitemOpen
  \bibfield  {author} {\bibinfo {author} {\bibfnamefont {J.~Y.}\ \bibnamefont
  {Liu}}, \bibinfo {author} {\bibfnamefont {J.}~\bibnamefont {Hu}}, \bibinfo
  {author} {\bibfnamefont {Q.}~\bibnamefont {Zhang}}, \bibinfo {author}
  {\bibfnamefont {D.}~\bibnamefont {Graf}}, \bibinfo {author} {\bibfnamefont
  {H.~B.}\ \bibnamefont {Cao}}, \bibinfo {author} {\bibfnamefont {S.~M.~A.}\
  \bibnamefont {Radmanesh}}, \bibinfo {author} {\bibfnamefont {D.~J.}\
  \bibnamefont {Adams}}, \bibinfo {author} {\bibfnamefont {Y.~L.}\ \bibnamefont
  {Zhu}}, \bibinfo {author} {\bibfnamefont {G.~.~F.}\ \bibnamefont {Cheng}},
  \bibinfo {author} {\bibfnamefont {X.}~\bibnamefont {Liu}}, \bibinfo {author}
  {\bibfnamefont {W.~A.}\ \bibnamefont {Phelan}}, \bibinfo {author}
  {\bibfnamefont {J.}~\bibnamefont {Wei}}, \bibinfo {author} {\bibfnamefont
  {M.}~\bibnamefont {Jaime}}, \bibinfo {author} {\bibfnamefont
  {F.}~\bibnamefont {Balakirev}}, \bibinfo {author} {\bibfnamefont {D.~A.}\
  \bibnamefont {Tennant}}, \bibinfo {author} {\bibfnamefont {J.~F.}\
  \bibnamefont {DiTusa}}, \bibinfo {author} {\bibfnamefont {I.}~\bibnamefont
  {Chiorescu}}, \bibinfo {author} {\bibfnamefont {L.}~\bibnamefont {Spinu}}, \
  and\ \bibinfo {author} {\bibfnamefont {Z.~Q.}\ \bibnamefont {Mao}},\ }\href
  {\doibase 10.1038/nmat4953} {\bibfield  {journal} {\bibinfo  {journal}
  {Nature Materials}\ }\textbf {\bibinfo {volume} {16}},\ \bibinfo {pages}
  {905} (\bibinfo {year} {2017})}\BibitemShut {NoStop}%
\bibitem [{\citenamefont {Suzuki}\ \emph {et~al.}(2016)\citenamefont {Suzuki},
  \citenamefont {Chisnell}, \citenamefont {Devarakonda}, \citenamefont {Liu},
  \citenamefont {Feng}, \citenamefont {Xiao}, \citenamefont {Lynn},\ and\
  \citenamefont {Checkelsky}}]{GdPtBi}%
  \BibitemOpen
  \bibfield  {author} {\bibinfo {author} {\bibfnamefont {T.}~\bibnamefont
  {Suzuki}}, \bibinfo {author} {\bibfnamefont {R.}~\bibnamefont {Chisnell}},
  \bibinfo {author} {\bibfnamefont {A.}~\bibnamefont {Devarakonda}}, \bibinfo
  {author} {\bibfnamefont {Y.-T.}\ \bibnamefont {Liu}}, \bibinfo {author}
  {\bibfnamefont {W.}~\bibnamefont {Feng}}, \bibinfo {author} {\bibfnamefont
  {D.}~\bibnamefont {Xiao}}, \bibinfo {author} {\bibfnamefont {J.~.~W.}\
  \bibnamefont {Lynn}}, \ and\ \bibinfo {author} {\bibfnamefont {J.~.~G.}\
  \bibnamefont {Checkelsky}},\ }\href {\doibase 10.1038/nphys3831} {\bibfield
  {journal} {\bibinfo  {journal} {Nature Physics}\ }\textbf {\bibinfo {volume}
  {12}},\ \bibinfo {pages} {1119} (\bibinfo {year} {2016})}\BibitemShut
  {NoStop}%
\bibitem [{\citenamefont {Puphal}\ \emph {et~al.}(2020)\citenamefont {Puphal},
  \citenamefont {Pomjakushin}, \citenamefont {Kanazawa}, \citenamefont
  {Ukleev}, \citenamefont {Gawryluk}, \citenamefont {Ma}, \citenamefont
  {Naamneh}, \citenamefont {Plumb}, \citenamefont {Keller}, \citenamefont
  {Cubitt}, \citenamefont {Pomjakushina},\ and\ \citenamefont
  {White}}]{CeAlGe2020}%
  \BibitemOpen
  \bibfield  {author} {\bibinfo {author} {\bibfnamefont {P.}~\bibnamefont
  {Puphal}}, \bibinfo {author} {\bibfnamefont {V.}~\bibnamefont {Pomjakushin}},
  \bibinfo {author} {\bibfnamefont {N.}~\bibnamefont {Kanazawa}}, \bibinfo
  {author} {\bibfnamefont {V.}~\bibnamefont {Ukleev}}, \bibinfo {author}
  {\bibfnamefont {D.~J.}\ \bibnamefont {Gawryluk}}, \bibinfo {author}
  {\bibfnamefont {J.}~\bibnamefont {Ma}}, \bibinfo {author} {\bibfnamefont
  {M.}~\bibnamefont {Naamneh}}, \bibinfo {author} {\bibfnamefont {N.~C.}\
  \bibnamefont {Plumb}}, \bibinfo {author} {\bibfnamefont {L.}~\bibnamefont
  {Keller}}, \bibinfo {author} {\bibfnamefont {R.}~\bibnamefont {Cubitt}},
  \bibinfo {author} {\bibfnamefont {E.}~\bibnamefont {Pomjakushina}}, \ and\
  \bibinfo {author} {\bibfnamefont {J.~S.}\ \bibnamefont {White}},\ }\href
  {\doibase 10.1103/PhysRevLett.124.017202} {\bibfield  {journal} {\bibinfo
  {journal} {Phys. Rev. Lett.}\ }\textbf {\bibinfo {volume} {124}},\ \bibinfo
  {pages} {017202} (\bibinfo {year} {2020})}\BibitemShut {NoStop}%
\bibitem [{\citenamefont {Bruno}\ \emph {et~al.}(2004)\citenamefont {Bruno},
  \citenamefont {Dugaev},\ and\ \citenamefont
  {Taillefumier}}]{PhysRevLett2004}%
  \BibitemOpen
  \bibfield  {author} {\bibinfo {author} {\bibfnamefont {P.}~\bibnamefont
  {Bruno}}, \bibinfo {author} {\bibfnamefont {V.~K.}\ \bibnamefont {Dugaev}}, \
  and\ \bibinfo {author} {\bibfnamefont {M.}~\bibnamefont {Taillefumier}},\
  }\href {\doibase 10.1103/PhysRevLett.93.096806} {\bibfield  {journal}
  {\bibinfo  {journal} {Phys. Rev. Lett.}\ }\textbf {\bibinfo {volume} {93}},\
  \bibinfo {pages} {096806} (\bibinfo {year} {2004})}\BibitemShut {NoStop}%
\bibitem [{\citenamefont {Wang}\ \emph {et~al.}(2019)\citenamefont {Wang},
  \citenamefont {Daniels}, \citenamefont {Liao}, \citenamefont {Zhao},
  \citenamefont {Wang}, \citenamefont {Koster}, \citenamefont {Rijnders},
  \citenamefont {Chang}, \citenamefont {Xiao},\ and\ \citenamefont
  {Wu}}]{Wang2019}%
  \BibitemOpen
  \bibfield  {author} {\bibinfo {author} {\bibfnamefont {W.}~\bibnamefont
  {Wang}}, \bibinfo {author} {\bibfnamefont {M.~W.}\ \bibnamefont {Daniels}},
  \bibinfo {author} {\bibfnamefont {Z.}~\bibnamefont {Liao}}, \bibinfo {author}
  {\bibfnamefont {Y.}~\bibnamefont {Zhao}}, \bibinfo {author} {\bibfnamefont
  {J.}~\bibnamefont {Wang}}, \bibinfo {author} {\bibfnamefont {G.}~\bibnamefont
  {Koster}}, \bibinfo {author} {\bibfnamefont {G.}~\bibnamefont {Rijnders}},
  \bibinfo {author} {\bibfnamefont {C.-Z.}\ \bibnamefont {Chang}}, \bibinfo
  {author} {\bibfnamefont {D.}~\bibnamefont {Xiao}}, \ and\ \bibinfo {author}
  {\bibfnamefont {W.}~\bibnamefont {Wu}},\ }\href {\doibase
  10.1038/s41563-019-0454-9} {\bibfield  {journal} {\bibinfo  {journal} {Nature
  Materials}\ }\textbf {\bibinfo {volume} {18}},\ \bibinfo {pages} {1054}
  (\bibinfo {year} {2019})}\BibitemShut {NoStop}%
\bibitem [{\citenamefont {Neubauer}\ \emph {et~al.}(2009)\citenamefont
  {Neubauer}, \citenamefont {Pfleiderer}, \citenamefont {Binz}, \citenamefont
  {Rosch}, \citenamefont {Ritz}, \citenamefont {Niklowitz},\ and\ \citenamefont
  {B\"oni}}]{MnSi}%
  \BibitemOpen
  \bibfield  {author} {\bibinfo {author} {\bibfnamefont {A.}~\bibnamefont
  {Neubauer}}, \bibinfo {author} {\bibfnamefont {C.}~\bibnamefont
  {Pfleiderer}}, \bibinfo {author} {\bibfnamefont {B.}~\bibnamefont {Binz}},
  \bibinfo {author} {\bibfnamefont {A.}~\bibnamefont {Rosch}}, \bibinfo
  {author} {\bibfnamefont {R.}~\bibnamefont {Ritz}}, \bibinfo {author}
  {\bibfnamefont {P.~G.}\ \bibnamefont {Niklowitz}}, \ and\ \bibinfo {author}
  {\bibfnamefont {P.}~\bibnamefont {B\"oni}},\ }\href {\doibase
  10.1103/PhysRevLett.102.186602} {\bibfield  {journal} {\bibinfo  {journal}
  {Phys. Rev. Lett.}\ }\textbf {\bibinfo {volume} {102}},\ \bibinfo {pages}
  {186602} (\bibinfo {year} {2009})}\BibitemShut {NoStop}%
\bibitem [{\citenamefont {Kanazawa}\ \emph {et~al.}(2011)\citenamefont
  {Kanazawa}, \citenamefont {Onose}, \citenamefont {Arima}, \citenamefont
  {Okuyama}, \citenamefont {Ohoyama}, \citenamefont {Wakimoto}, \citenamefont
  {Kakurai}, \citenamefont {Ishiwata},\ and\ \citenamefont {Tokura}}]{MnGe}%
  \BibitemOpen
  \bibfield  {author} {\bibinfo {author} {\bibfnamefont {N.}~\bibnamefont
  {Kanazawa}}, \bibinfo {author} {\bibfnamefont {Y.}~\bibnamefont {Onose}},
  \bibinfo {author} {\bibfnamefont {T.}~\bibnamefont {Arima}}, \bibinfo
  {author} {\bibfnamefont {D.}~\bibnamefont {Okuyama}}, \bibinfo {author}
  {\bibfnamefont {K.}~\bibnamefont {Ohoyama}}, \bibinfo {author} {\bibfnamefont
  {S.}~\bibnamefont {Wakimoto}}, \bibinfo {author} {\bibfnamefont
  {K.}~\bibnamefont {Kakurai}}, \bibinfo {author} {\bibfnamefont
  {S.}~\bibnamefont {Ishiwata}}, \ and\ \bibinfo {author} {\bibfnamefont
  {Y.}~\bibnamefont {Tokura}},\ }\href {\doibase
  10.1103/PhysRevLett.106.156603} {\bibfield  {journal} {\bibinfo  {journal}
  {Phys. Rev. Lett.}\ }\textbf {\bibinfo {volume} {106}},\ \bibinfo {pages}
  {156603} (\bibinfo {year} {2011})}\BibitemShut {NoStop}%
\bibitem [{\citenamefont {S{\"u}rgers}\ \emph {et~al.}(2014)\citenamefont
  {S{\"u}rgers}, \citenamefont {Fischer}, \citenamefont {Winkel},\ and\
  \citenamefont {L{\"o}hneysen}}]{Surgers2014}%
  \BibitemOpen
  \bibfield  {author} {\bibinfo {author} {\bibfnamefont {C.}~\bibnamefont
  {S{\"u}rgers}}, \bibinfo {author} {\bibfnamefont {G.}~\bibnamefont
  {Fischer}}, \bibinfo {author} {\bibfnamefont {P.}~\bibnamefont {Winkel}}, \
  and\ \bibinfo {author} {\bibfnamefont {H.~v.}\ \bibnamefont
  {L{\"o}hneysen}},\ }\href {\doibase 10.1038/ncomms4400} {\bibfield  {journal}
  {\bibinfo  {journal} {Nature Communications}\ }\textbf {\bibinfo {volume}
  {5}},\ \bibinfo {pages} {3400} (\bibinfo {year} {2014})}\BibitemShut
  {NoStop}%
\bibitem [{\citenamefont {Vistoli}\ \emph {et~al.}(2019)\citenamefont
  {Vistoli}, \citenamefont {Wang}, \citenamefont {Sander}, \citenamefont {Zhu},
  \citenamefont {Casals}, \citenamefont {Cichelero}, \citenamefont
  {Barth{\'e}l{\'e}my}, \citenamefont {Fusil}, \citenamefont {Herranz},
  \citenamefont {Valencia}, \citenamefont {Abrudan}, \citenamefont {Weschke},
  \citenamefont {Nakazawa}, \citenamefont {Kohno}, \citenamefont {Santamaria},
  \citenamefont {Wu}, \citenamefont {Garcia},\ and\ \citenamefont
  {Bibes}}]{Vistoli2019}%
  \BibitemOpen
  \bibfield  {author} {\bibinfo {author} {\bibfnamefont {L.}~\bibnamefont
  {Vistoli}}, \bibinfo {author} {\bibfnamefont {W.}~\bibnamefont {Wang}},
  \bibinfo {author} {\bibfnamefont {A.}~\bibnamefont {Sander}}, \bibinfo
  {author} {\bibfnamefont {Q.}~\bibnamefont {Zhu}}, \bibinfo {author}
  {\bibfnamefont {B.}~\bibnamefont {Casals}}, \bibinfo {author} {\bibfnamefont
  {R.}~\bibnamefont {Cichelero}}, \bibinfo {author} {\bibfnamefont
  {A.}~\bibnamefont {Barth{\'e}l{\'e}my}}, \bibinfo {author} {\bibfnamefont
  {S.}~\bibnamefont {Fusil}}, \bibinfo {author} {\bibfnamefont
  {G.}~\bibnamefont {Herranz}}, \bibinfo {author} {\bibfnamefont
  {S.}~\bibnamefont {Valencia}}, \bibinfo {author} {\bibfnamefont
  {R.}~\bibnamefont {Abrudan}}, \bibinfo {author} {\bibfnamefont
  {E.}~\bibnamefont {Weschke}}, \bibinfo {author} {\bibfnamefont
  {K.}~\bibnamefont {Nakazawa}}, \bibinfo {author} {\bibfnamefont
  {H.}~\bibnamefont {Kohno}}, \bibinfo {author} {\bibfnamefont
  {J.}~\bibnamefont {Santamaria}}, \bibinfo {author} {\bibfnamefont
  {W.}~\bibnamefont {Wu}}, \bibinfo {author} {\bibfnamefont {V.}~\bibnamefont
  {Garcia}}, \ and\ \bibinfo {author} {\bibfnamefont {M.}~\bibnamefont
  {Bibes}},\ }\href {\doibase 10.1038/s41567-018-0307-5} {\bibfield  {journal}
  {\bibinfo  {journal} {Nature Physics}\ }\textbf {\bibinfo {volume} {15}},\
  \bibinfo {pages} {67} (\bibinfo {year} {2019})}\BibitemShut {NoStop}%
\bibitem [{\citenamefont {Laha}\ \emph {et~al.}(2021)\citenamefont {Laha},
  \citenamefont {Singha}, \citenamefont {Mardanya}, \citenamefont {Singh},
  \citenamefont {Agarwal}, \citenamefont {Mandal},\ and\ \citenamefont
  {Hossain}}]{EuAgAs}%
  \BibitemOpen
  \bibfield  {author} {\bibinfo {author} {\bibfnamefont {A.}~\bibnamefont
  {Laha}}, \bibinfo {author} {\bibfnamefont {R.}~\bibnamefont {Singha}},
  \bibinfo {author} {\bibfnamefont {S.}~\bibnamefont {Mardanya}}, \bibinfo
  {author} {\bibfnamefont {B.}~\bibnamefont {Singh}}, \bibinfo {author}
  {\bibfnamefont {A.}~\bibnamefont {Agarwal}}, \bibinfo {author} {\bibfnamefont
  {P.}~\bibnamefont {Mandal}}, \ and\ \bibinfo {author} {\bibfnamefont
  {Z.}~\bibnamefont {Hossain}},\ }\href {\doibase 10.1103/PhysRevB.103.L241112}
  {\bibfield  {journal} {\bibinfo  {journal} {Phys. Rev. B}\ }\textbf {\bibinfo
  {volume} {103}},\ \bibinfo {pages} {L241112} (\bibinfo {year}
  {2021})}\BibitemShut {NoStop}%
\bibitem [{\citenamefont {Ohuchi}\ \emph {et~al.}(2015)\citenamefont {Ohuchi},
  \citenamefont {Kozuka}, \citenamefont {Uchida}, \citenamefont {Ueno},
  \citenamefont {Tsukazaki},\ and\ \citenamefont {Kawasaki}}]{EuO}%
  \BibitemOpen
  \bibfield  {author} {\bibinfo {author} {\bibfnamefont {Y.}~\bibnamefont
  {Ohuchi}}, \bibinfo {author} {\bibfnamefont {Y.}~\bibnamefont {Kozuka}},
  \bibinfo {author} {\bibfnamefont {M.}~\bibnamefont {Uchida}}, \bibinfo
  {author} {\bibfnamefont {K.}~\bibnamefont {Ueno}}, \bibinfo {author}
  {\bibfnamefont {A.}~\bibnamefont {Tsukazaki}}, \ and\ \bibinfo {author}
  {\bibfnamefont {M.}~\bibnamefont {Kawasaki}},\ }\href {\doibase
  10.1103/PhysRevB.91.245115} {\bibfield  {journal} {\bibinfo  {journal} {Phys.
  Rev. B}\ }\textbf {\bibinfo {volume} {91}},\ \bibinfo {pages} {245115}
  (\bibinfo {year} {2015})}\BibitemShut {NoStop}%
\bibitem [{\citenamefont {Machida}\ \emph {et~al.}(2007)\citenamefont
  {Machida}, \citenamefont {Nakatsuji}, \citenamefont {Maeno}, \citenamefont
  {Tayama}, \citenamefont {Sakakibara},\ and\ \citenamefont
  {Onoda}}]{Pr2Ir2O7}%
  \BibitemOpen
  \bibfield  {author} {\bibinfo {author} {\bibfnamefont {Y.}~\bibnamefont
  {Machida}}, \bibinfo {author} {\bibfnamefont {S.}~\bibnamefont {Nakatsuji}},
  \bibinfo {author} {\bibfnamefont {Y.}~\bibnamefont {Maeno}}, \bibinfo
  {author} {\bibfnamefont {T.}~\bibnamefont {Tayama}}, \bibinfo {author}
  {\bibfnamefont {T.}~\bibnamefont {Sakakibara}}, \ and\ \bibinfo {author}
  {\bibfnamefont {S.}~\bibnamefont {Onoda}},\ }\href {\doibase
  10.1103/PhysRevLett.98.057203} {\bibfield  {journal} {\bibinfo  {journal}
  {Phys. Rev. Lett.}\ }\textbf {\bibinfo {volume} {98}},\ \bibinfo {pages}
  {057203} (\bibinfo {year} {2007})}\BibitemShut {NoStop}%
\bibitem [{\citenamefont {Pöttgen}\ \emph {et~al.}(1998)\citenamefont
  {Pöttgen}, \citenamefont {Kotzyba},\ and\ \citenamefont
  {Görlich}}]{GdAgGe}%
  \BibitemOpen
  \bibfield  {author} {\bibinfo {author} {\bibfnamefont {R.}~\bibnamefont
  {Pöttgen}}, \bibinfo {author} {\bibfnamefont {G.}~\bibnamefont {Kotzyba}}, \
  and\ \bibinfo {author} {\bibfnamefont {E.~A.}\ \bibnamefont {Görlich}},\
  }\href {https://www.sciencedirect.com/science/article/pii/S0022459698979476}
  {\bibfield  {journal} {\bibinfo  {journal} {Journal of Solid State
  Chemistry}\ }\textbf {\bibinfo {volume} {141}},\ \bibinfo {pages} {352}
  (\bibinfo {year} {1998})}\BibitemShut {NoStop}%
\bibitem [{\citenamefont {Baran}\ \emph {et~al.}(1998)\citenamefont {Baran},
  \citenamefont {Hofmann}, \citenamefont {Leciejewicz}, \citenamefont {Penc},
  \citenamefont {Ślaski},\ and\ \citenamefont {Szytuła}}]{RAgGe_1998}%
  \BibitemOpen
  \bibfield  {author} {\bibinfo {author} {\bibfnamefont {S.}~\bibnamefont
  {Baran}}, \bibinfo {author} {\bibfnamefont {M.}~\bibnamefont {Hofmann}},
  \bibinfo {author} {\bibfnamefont {J.}~\bibnamefont {Leciejewicz}}, \bibinfo
  {author} {\bibfnamefont {B.}~\bibnamefont {Penc}}, \bibinfo {author}
  {\bibfnamefont {M.}~\bibnamefont {Ślaski}}, \ and\ \bibinfo {author}
  {\bibfnamefont {A.}~\bibnamefont {Szytuła}},\ }\href {\doibase
  https://doi.org/10.1016/S0925-8388(98)00721-X} {\bibfield  {journal}
  {\bibinfo  {journal} {Journal of Alloys and Compounds}\ }\textbf {\bibinfo
  {volume} {281}},\ \bibinfo {pages} {92} (\bibinfo {year} {1998})}\BibitemShut
  {NoStop}%
\bibitem [{\citenamefont {Morosan}\ \emph {et~al.}(2004)\citenamefont
  {Morosan}, \citenamefont {Bud’ko}, \citenamefont {Canfield}, \citenamefont
  {Torikachvili},\ and\ \citenamefont {Lacerda}}]{RAgGe2004}%
  \BibitemOpen
  \bibfield  {author} {\bibinfo {author} {\bibfnamefont {E.}~\bibnamefont
  {Morosan}}, \bibinfo {author} {\bibfnamefont {S.}~\bibnamefont {Bud’ko}},
  \bibinfo {author} {\bibfnamefont {P.}~\bibnamefont {Canfield}}, \bibinfo
  {author} {\bibfnamefont {M.}~\bibnamefont {Torikachvili}}, \ and\ \bibinfo
  {author} {\bibfnamefont {A.}~\bibnamefont {Lacerda}},\ }\href {\doibase
  https://doi.org/10.1016/j.jmmm.2003.11.014} {\bibfield  {journal} {\bibinfo
  {journal} {Journal of Magnetism and Magnetic Materials}\ }\textbf {\bibinfo
  {volume} {277}},\ \bibinfo {pages} {298} (\bibinfo {year}
  {2004})}\BibitemShut {NoStop}%
\bibitem [{\citenamefont {Goddard}\ \emph {et~al.}(2007)\citenamefont
  {Goddard}, \citenamefont {Singleton}, \citenamefont {Lima~Sharma},
  \citenamefont {Morosan}, \citenamefont {Blundell}, \citenamefont {Bud'ko},\
  and\ \citenamefont {Canfield}}]{TmAgGe}%
  \BibitemOpen
  \bibfield  {author} {\bibinfo {author} {\bibfnamefont {P.~A.}\ \bibnamefont
  {Goddard}}, \bibinfo {author} {\bibfnamefont {J.}~\bibnamefont {Singleton}},
  \bibinfo {author} {\bibfnamefont {A.~L.}\ \bibnamefont {Lima~Sharma}},
  \bibinfo {author} {\bibfnamefont {E.}~\bibnamefont {Morosan}}, \bibinfo
  {author} {\bibfnamefont {S.~J.}\ \bibnamefont {Blundell}}, \bibinfo {author}
  {\bibfnamefont {S.~L.}\ \bibnamefont {Bud'ko}}, \ and\ \bibinfo {author}
  {\bibfnamefont {P.~C.}\ \bibnamefont {Canfield}},\ }\href {\doibase
  10.1103/PhysRevB.75.094426} {\bibfield  {journal} {\bibinfo  {journal} {Phys.
  Rev. B}\ }\textbf {\bibinfo {volume} {75}},\ \bibinfo {pages} {094426}
  (\bibinfo {year} {2007})}\BibitemShut {NoStop}%
\bibitem [{\citenamefont {Chang}\ \emph {et~al.}(2018)\citenamefont {Chang},
  \citenamefont {Singh}, \citenamefont {Xu}, \citenamefont {Bian},
  \citenamefont {Huang}, \citenamefont {Hsu}, \citenamefont {Belopolski},
  \citenamefont {Alidoust}, \citenamefont {Sanchez}, \citenamefont {Zheng},
  \citenamefont {Lu}, \citenamefont {Zhang}, \citenamefont {Bian},
  \citenamefont {Chang}, \citenamefont {Jeng}, \citenamefont {Bansil},
  \citenamefont {Hsu}, \citenamefont {Jia}, \citenamefont {Neupert},
  \citenamefont {Lin},\ and\ \citenamefont {Hasan}}]{RAlGe}%
  \BibitemOpen
  \bibfield  {author} {\bibinfo {author} {\bibfnamefont {G.}~\bibnamefont
  {Chang}}, \bibinfo {author} {\bibfnamefont {B.}~\bibnamefont {Singh}},
  \bibinfo {author} {\bibfnamefont {S.-Y.}\ \bibnamefont {Xu}}, \bibinfo
  {author} {\bibfnamefont {G.}~\bibnamefont {Bian}}, \bibinfo {author}
  {\bibfnamefont {S.-M.}\ \bibnamefont {Huang}}, \bibinfo {author}
  {\bibfnamefont {C.-H.}\ \bibnamefont {Hsu}}, \bibinfo {author} {\bibfnamefont
  {I.}~\bibnamefont {Belopolski}}, \bibinfo {author} {\bibfnamefont
  {N.}~\bibnamefont {Alidoust}}, \bibinfo {author} {\bibfnamefont {D.~S.}\
  \bibnamefont {Sanchez}}, \bibinfo {author} {\bibfnamefont {H.}~\bibnamefont
  {Zheng}}, \bibinfo {author} {\bibfnamefont {H.}~\bibnamefont {Lu}}, \bibinfo
  {author} {\bibfnamefont {X.}~\bibnamefont {Zhang}}, \bibinfo {author}
  {\bibfnamefont {Y.}~\bibnamefont {Bian}}, \bibinfo {author} {\bibfnamefont
  {T.-R.}\ \bibnamefont {Chang}}, \bibinfo {author} {\bibfnamefont {H.-T.}\
  \bibnamefont {Jeng}}, \bibinfo {author} {\bibfnamefont {A.}~\bibnamefont
  {Bansil}}, \bibinfo {author} {\bibfnamefont {H.}~\bibnamefont {Hsu}},
  \bibinfo {author} {\bibfnamefont {S.}~\bibnamefont {Jia}}, \bibinfo {author}
  {\bibfnamefont {T.}~\bibnamefont {Neupert}}, \bibinfo {author} {\bibfnamefont
  {H.}~\bibnamefont {Lin}}, \ and\ \bibinfo {author} {\bibfnamefont {M.~Z.}\
  \bibnamefont {Hasan}},\ }\href {\doibase 10.1103/PhysRevB.97.041104}
  {\bibfield  {journal} {\bibinfo  {journal} {Phys. Rev. B}\ }\textbf {\bibinfo
  {volume} {97}},\ \bibinfo {pages} {041104} (\bibinfo {year}
  {2018})}\BibitemShut {NoStop}%
\bibitem [{\citenamefont {Destraz}\ \emph {et~al.}(2020)\citenamefont
  {Destraz}, \citenamefont {Das}, \citenamefont {Tsirkin}, \citenamefont {Xu},
  \citenamefont {Neupert}, \citenamefont {Chang}, \citenamefont {Schilling},
  \citenamefont {Grushin}, \citenamefont {Kohlbrecher}, \citenamefont {Keller},
  \citenamefont {Puphal}, \citenamefont {Pomjakushina},\ and\ \citenamefont
  {White}}]{PrAlGe2020}%
  \BibitemOpen
  \bibfield  {author} {\bibinfo {author} {\bibfnamefont {D.}~\bibnamefont
  {Destraz}}, \bibinfo {author} {\bibfnamefont {L.}~\bibnamefont {Das}},
  \bibinfo {author} {\bibfnamefont {S.~S.}\ \bibnamefont {Tsirkin}}, \bibinfo
  {author} {\bibfnamefont {Y.}~\bibnamefont {Xu}}, \bibinfo {author}
  {\bibfnamefont {T.}~\bibnamefont {Neupert}}, \bibinfo {author} {\bibfnamefont
  {J.}~\bibnamefont {Chang}}, \bibinfo {author} {\bibfnamefont
  {A.}~\bibnamefont {Schilling}}, \bibinfo {author} {\bibfnamefont {A.~G.}\
  \bibnamefont {Grushin}}, \bibinfo {author} {\bibfnamefont {J.}~\bibnamefont
  {Kohlbrecher}}, \bibinfo {author} {\bibfnamefont {L.}~\bibnamefont {Keller}},
  \bibinfo {author} {\bibfnamefont {P.}~\bibnamefont {Puphal}}, \bibinfo
  {author} {\bibfnamefont {E.}~\bibnamefont {Pomjakushina}}, \ and\ \bibinfo
  {author} {\bibfnamefont {J.~S.}\ \bibnamefont {White}},\ }\href {\doibase
  10.1038/s41535-019-0207-7} {\bibfield  {journal} {\bibinfo  {journal} {npj
  Quantum Materials}\ }\textbf {\bibinfo {volume} {5}},\ \bibinfo {pages} {5}
  (\bibinfo {year} {2020})}\BibitemShut {NoStop}%
\bibitem [{\citenamefont {Sanchez}\ \emph {et~al.}(2020)\citenamefont
  {Sanchez}, \citenamefont {Chang}, \citenamefont {Belopolski}, \citenamefont
  {Lu}, \citenamefont {Yin}, \citenamefont {Alidoust}, \citenamefont {Xu},
  \citenamefont {Cochran}, \citenamefont {Zhang}, \citenamefont {Bian},
  \citenamefont {Zhang}, \citenamefont {Liu}, \citenamefont {Ma}, \citenamefont
  {Bian}, \citenamefont {Lin}, \citenamefont {Xu}, \citenamefont {Jia},\ and\
  \citenamefont {Hasan}}]{PrAlGe2020_2}%
  \BibitemOpen
  \bibfield  {author} {\bibinfo {author} {\bibfnamefont {D.~S.}\ \bibnamefont
  {Sanchez}}, \bibinfo {author} {\bibfnamefont {G.}~\bibnamefont {Chang}},
  \bibinfo {author} {\bibfnamefont {I.}~\bibnamefont {Belopolski}}, \bibinfo
  {author} {\bibfnamefont {H.}~\bibnamefont {Lu}}, \bibinfo {author}
  {\bibfnamefont {J.-X.}\ \bibnamefont {Yin}}, \bibinfo {author} {\bibfnamefont
  {N.}~\bibnamefont {Alidoust}}, \bibinfo {author} {\bibfnamefont
  {X.}~\bibnamefont {Xu}}, \bibinfo {author} {\bibfnamefont {T.~A.}\
  \bibnamefont {Cochran}}, \bibinfo {author} {\bibfnamefont {X.}~\bibnamefont
  {Zhang}}, \bibinfo {author} {\bibfnamefont {Y.}~\bibnamefont {Bian}},
  \bibinfo {author} {\bibfnamefont {S.~S.}\ \bibnamefont {Zhang}}, \bibinfo
  {author} {\bibfnamefont {Y.-Y.}\ \bibnamefont {Liu}}, \bibinfo {author}
  {\bibfnamefont {J.}~\bibnamefont {Ma}}, \bibinfo {author} {\bibfnamefont
  {G.}~\bibnamefont {Bian}}, \bibinfo {author} {\bibfnamefont {H.}~\bibnamefont
  {Lin}}, \bibinfo {author} {\bibfnamefont {S.-Y.}\ \bibnamefont {Xu}},
  \bibinfo {author} {\bibfnamefont {S.}~\bibnamefont {Jia}}, \ and\ \bibinfo
  {author} {\bibfnamefont {M.~Z.}\ \bibnamefont {Hasan}},\ }\href {\doibase
  10.1038/s41467-020-16879-1} {\bibfield  {journal} {\bibinfo  {journal}
  {Nature Communications}\ }\textbf {\bibinfo {volume} {11}},\ \bibinfo {pages}
  {3356} (\bibinfo {year} {2020})}\BibitemShut {NoStop}%
\bibitem [{\citenamefont {Hodovanets}\ \emph {et~al.}(2018)\citenamefont
  {Hodovanets}, \citenamefont {Eckberg}, \citenamefont {Zavalij}, \citenamefont
  {Kim}, \citenamefont {Lin}, \citenamefont {Zic}, \citenamefont {Campbell},
  \citenamefont {Higgins},\ and\ \citenamefont {Paglione}}]{CeAlGe2018}%
  \BibitemOpen
  \bibfield  {author} {\bibinfo {author} {\bibfnamefont {H.}~\bibnamefont
  {Hodovanets}}, \bibinfo {author} {\bibfnamefont {C.~J.}\ \bibnamefont
  {Eckberg}}, \bibinfo {author} {\bibfnamefont {P.~Y.}\ \bibnamefont
  {Zavalij}}, \bibinfo {author} {\bibfnamefont {H.}~\bibnamefont {Kim}},
  \bibinfo {author} {\bibfnamefont {W.-C.}\ \bibnamefont {Lin}}, \bibinfo
  {author} {\bibfnamefont {M.}~\bibnamefont {Zic}}, \bibinfo {author}
  {\bibfnamefont {D.~J.}\ \bibnamefont {Campbell}}, \bibinfo {author}
  {\bibfnamefont {J.~S.}\ \bibnamefont {Higgins}}, \ and\ \bibinfo {author}
  {\bibfnamefont {J.}~\bibnamefont {Paglione}},\ }\href {\doibase
  10.1103/PhysRevB.98.245132} {\bibfield  {journal} {\bibinfo  {journal} {Phys.
  Rev. B}\ }\textbf {\bibinfo {volume} {98}},\ \bibinfo {pages} {245132}
  (\bibinfo {year} {2018})}\BibitemShut {NoStop}%
\bibitem [{\citenamefont {Meng}\ \emph {et~al.}(2019)\citenamefont {Meng},
  \citenamefont {Wu}, \citenamefont {Qiu}, \citenamefont {Wang}, \citenamefont
  {Liu}, \citenamefont {Xia}, \citenamefont {Yuan}, \citenamefont {Chang},\
  and\ \citenamefont {Tian}}]{PrAlGe}%
  \BibitemOpen
  \bibfield  {author} {\bibinfo {author} {\bibfnamefont {B.}~\bibnamefont
  {Meng}}, \bibinfo {author} {\bibfnamefont {H.}~\bibnamefont {Wu}}, \bibinfo
  {author} {\bibfnamefont {Y.}~\bibnamefont {Qiu}}, \bibinfo {author}
  {\bibfnamefont {C.}~\bibnamefont {Wang}}, \bibinfo {author} {\bibfnamefont
  {Y.}~\bibnamefont {Liu}}, \bibinfo {author} {\bibfnamefont {Z.}~\bibnamefont
  {Xia}}, \bibinfo {author} {\bibfnamefont {S.}~\bibnamefont {Yuan}}, \bibinfo
  {author} {\bibfnamefont {H.}~\bibnamefont {Chang}}, \ and\ \bibinfo {author}
  {\bibfnamefont {Z.}~\bibnamefont {Tian}},\ }\href {\doibase
  10.1063/1.5090795} {\bibfield  {journal} {\bibinfo  {journal} {APL
  Materials}\ }\textbf {\bibinfo {volume} {7}},\ \bibinfo {pages} {051110}
  (\bibinfo {year} {2019})}\BibitemShut {NoStop}%
\bibitem [{\citenamefont {Bud'ko}\ \emph {et~al.}(2004)\citenamefont {Bud'ko},
  \citenamefont {Morosan},\ and\ \citenamefont {Canfield}}]{YbAgGe2004}%
  \BibitemOpen
  \bibfield  {author} {\bibinfo {author} {\bibfnamefont {S.~L.}\ \bibnamefont
  {Bud'ko}}, \bibinfo {author} {\bibfnamefont {E.}~\bibnamefont {Morosan}}, \
  and\ \bibinfo {author} {\bibfnamefont {P.~C.}\ \bibnamefont {Canfield}},\
  }\href {\doibase 10.1103/PhysRevB.69.014415} {\bibfield  {journal} {\bibinfo
  {journal} {Phys. Rev. B}\ }\textbf {\bibinfo {volume} {69}},\ \bibinfo
  {pages} {014415} (\bibinfo {year} {2004})}\BibitemShut {NoStop}%
\bibitem [{\citenamefont {Bud'ko}\ \emph {et~al.}(2005)\citenamefont {Bud'ko},
  \citenamefont {Morosan},\ and\ \citenamefont {Canfield}}]{YbAgGe2005}%
  \BibitemOpen
  \bibfield  {author} {\bibinfo {author} {\bibfnamefont {S.~L.}\ \bibnamefont
  {Bud'ko}}, \bibinfo {author} {\bibfnamefont {E.}~\bibnamefont {Morosan}}, \
  and\ \bibinfo {author} {\bibfnamefont {P.~C.}\ \bibnamefont {Canfield}},\
  }\href {\doibase 10.1103/PhysRevB.71.054408} {\bibfield  {journal} {\bibinfo
  {journal} {Phys. Rev. B}\ }\textbf {\bibinfo {volume} {71}},\ \bibinfo
  {pages} {054408} (\bibinfo {year} {2005})}\BibitemShut {NoStop}%
\bibitem [{\citenamefont {Tokiwa}\ \emph {et~al.}(2013)\citenamefont {Tokiwa},
  \citenamefont {Garst}, \citenamefont {Gegenwart}, \citenamefont {Bud'ko},\
  and\ \citenamefont {Canfield}}]{YbAgGe2013}%
  \BibitemOpen
  \bibfield  {author} {\bibinfo {author} {\bibfnamefont {Y.}~\bibnamefont
  {Tokiwa}}, \bibinfo {author} {\bibfnamefont {M.}~\bibnamefont {Garst}},
  \bibinfo {author} {\bibfnamefont {P.}~\bibnamefont {Gegenwart}}, \bibinfo
  {author} {\bibfnamefont {S.~L.}\ \bibnamefont {Bud'ko}}, \ and\ \bibinfo
  {author} {\bibfnamefont {P.~C.}\ \bibnamefont {Canfield}},\ }\href {\doibase
  10.1103/PhysRevLett.111.116401} {\bibfield  {journal} {\bibinfo  {journal}
  {Phys. Rev. Lett.}\ }\textbf {\bibinfo {volume} {111}},\ \bibinfo {pages}
  {116401} (\bibinfo {year} {2013})}\BibitemShut {NoStop}%
\bibitem [{\citenamefont {Hohenberg}\ and\ \citenamefont
  {Kohn}(1964)}]{Hohenberg}%
  \BibitemOpen
  \bibfield  {author} {\bibinfo {author} {\bibfnamefont {P.}~\bibnamefont
  {Hohenberg}}\ and\ \bibinfo {author} {\bibfnamefont {W.}~\bibnamefont
  {Kohn}},\ }\href {\doibase 10.1103/PhysRev.136.B864} {\bibfield  {journal}
  {\bibinfo  {journal} {Phys. Rev.}\ }\textbf {\bibinfo {volume} {136}},\
  \bibinfo {pages} {B864} (\bibinfo {year} {1964})}\BibitemShut {NoStop}%
\bibitem [{\citenamefont {Kohn}\ and\ \citenamefont {Sham}(1965)}]{Kohn}%
  \BibitemOpen
  \bibfield  {author} {\bibinfo {author} {\bibfnamefont {W.}~\bibnamefont
  {Kohn}}\ and\ \bibinfo {author} {\bibfnamefont {L.~J.}\ \bibnamefont
  {Sham}},\ }\href {\doibase 10.1103/PhysRev.140.A1133} {\bibfield  {journal}
  {\bibinfo  {journal} {Phys. Rev.}\ }\textbf {\bibinfo {volume} {140}},\
  \bibinfo {pages} {A1133} (\bibinfo {year} {1965})}\BibitemShut {NoStop}%
\bibitem [{\citenamefont {Bl\"ochl}(1994)}]{PAW}%
  \BibitemOpen
  \bibfield  {author} {\bibinfo {author} {\bibfnamefont {P.~E.}\ \bibnamefont
  {Bl\"ochl}},\ }\href {\doibase 10.1103/PhysRevB.50.17953} {\bibfield
  {journal} {\bibinfo  {journal} {Phys. Rev. B}\ }\textbf {\bibinfo {volume}
  {50}},\ \bibinfo {pages} {17953} (\bibinfo {year} {1994})}\BibitemShut
  {NoStop}%
\bibitem [{\citenamefont {Kresse}\ and\ \citenamefont
  {Furthm\"uller}(1996)}]{Kresse1}%
  \BibitemOpen
  \bibfield  {author} {\bibinfo {author} {\bibfnamefont {G.}~\bibnamefont
  {Kresse}}\ and\ \bibinfo {author} {\bibfnamefont {J.}~\bibnamefont
  {Furthm\"uller}},\ }\href {\doibase 10.1103/PhysRevB.54.11169} {\bibfield
  {journal} {\bibinfo  {journal} {Phys. Rev. B}\ }\textbf {\bibinfo {volume}
  {54}},\ \bibinfo {pages} {11169} (\bibinfo {year} {1996})}\BibitemShut
  {NoStop}%
\bibitem [{\citenamefont {Kresse}\ and\ \citenamefont
  {Joubert}(1999)}]{Kresse2}%
  \BibitemOpen
  \bibfield  {author} {\bibinfo {author} {\bibfnamefont {G.}~\bibnamefont
  {Kresse}}\ and\ \bibinfo {author} {\bibfnamefont {D.}~\bibnamefont
  {Joubert}},\ }\href {\doibase 10.1103/PhysRevB.59.1758} {\bibfield  {journal}
  {\bibinfo  {journal} {Phys. Rev. B}\ }\textbf {\bibinfo {volume} {59}},\
  \bibinfo {pages} {1758} (\bibinfo {year} {1999})}\BibitemShut {NoStop}%
\bibitem [{\citenamefont {Perdew}\ \emph {et~al.}(1996)\citenamefont {Perdew},
  \citenamefont {Burke},\ and\ \citenamefont {Ernzerhof}}]{PBE}%
  \BibitemOpen
  \bibfield  {author} {\bibinfo {author} {\bibfnamefont {J.~P.}\ \bibnamefont
  {Perdew}}, \bibinfo {author} {\bibfnamefont {K.}~\bibnamefont {Burke}}, \
  and\ \bibinfo {author} {\bibfnamefont {M.}~\bibnamefont {Ernzerhof}},\ }\href
  {\doibase 10.1103/PhysRevLett.77.3865} {\bibfield  {journal} {\bibinfo
  {journal} {Phys. Rev. Lett.}\ }\textbf {\bibinfo {volume} {77}},\ \bibinfo
  {pages} {3865} (\bibinfo {year} {1996})}\BibitemShut {NoStop}%
\bibitem [{\citenamefont {Petersen}\ \emph {et~al.}(2006)\citenamefont
  {Petersen}, \citenamefont {Hafner},\ and\ \citenamefont
  {Marsman}}]{Hubbard1}%
  \BibitemOpen
  \bibfield  {author} {\bibinfo {author} {\bibfnamefont {M.}~\bibnamefont
  {Petersen}}, \bibinfo {author} {\bibfnamefont {J.}~\bibnamefont {Hafner}}, \
  and\ \bibinfo {author} {\bibfnamefont {M.}~\bibnamefont {Marsman}},\ }\href
  {\doibase 10.1088/0953-8984/18/30/007} {\bibfield  {journal} {\bibinfo
  {journal} {Journal of Physics: Condensed Matter}\ }\textbf {\bibinfo {volume}
  {18}},\ \bibinfo {pages} {7021} (\bibinfo {year} {2006})}\BibitemShut
  {NoStop}%
\bibitem [{\citenamefont {Li}\ \emph {et~al.}(2015)\citenamefont {Li},
  \citenamefont {Su}, \citenamefont {Yang},\ and\ \citenamefont
  {Zhang}}]{Hubbard2}%
  \BibitemOpen
  \bibfield  {author} {\bibinfo {author} {\bibfnamefont {Z.}~\bibnamefont
  {Li}}, \bibinfo {author} {\bibfnamefont {H.}~\bibnamefont {Su}}, \bibinfo
  {author} {\bibfnamefont {X.}~\bibnamefont {Yang}}, \ and\ \bibinfo {author}
  {\bibfnamefont {J.}~\bibnamefont {Zhang}},\ }\href {\doibase
  10.1103/PhysRevB.91.235128} {\bibfield  {journal} {\bibinfo  {journal} {Phys.
  Rev. B}\ }\textbf {\bibinfo {volume} {91}},\ \bibinfo {pages} {235128}
  (\bibinfo {year} {2015})}\BibitemShut {NoStop}%
\bibitem [{\citenamefont {Monkhorst}\ and\ \citenamefont
  {Pack}(1976)}]{Monkhorst}%
  \BibitemOpen
  \bibfield  {author} {\bibinfo {author} {\bibfnamefont {H.~J.}\ \bibnamefont
  {Monkhorst}}\ and\ \bibinfo {author} {\bibfnamefont {J.~D.}\ \bibnamefont
  {Pack}},\ }\href {\doibase 10.1103/PhysRevB.13.5188} {\bibfield  {journal}
  {\bibinfo  {journal} {Phys. Rev. B}\ }\textbf {\bibinfo {volume} {13}},\
  \bibinfo {pages} {5188} (\bibinfo {year} {1976})}\BibitemShut {NoStop}%
\bibitem [{\citenamefont {Pizzi}\ \emph {et~al.}(2020)\citenamefont {Pizzi},
  \citenamefont {Vitale}, \citenamefont {Arita}, \citenamefont {Blügel},
  \citenamefont {Freimuth}, \citenamefont {G{\'{e}}ranton}, \citenamefont
  {Gibertini}, \citenamefont {Gresch}, \citenamefont {Johnson}, \citenamefont
  {Koretsune}, \citenamefont {Iba{\~{n}}ez-Azpiroz}, \citenamefont {Lee},
  \citenamefont {Lihm}, \citenamefont {Marchand}, \citenamefont {Marrazzo},
  \citenamefont {Mokrousov}, \citenamefont {Mustafa}, \citenamefont {Nohara},
  \citenamefont {Nomura}, \citenamefont {Paulatto}, \citenamefont
  {Ponc{\'{e}}}, \citenamefont {Ponweiser}, \citenamefont {Qiao}, \citenamefont
  {Thöle}, \citenamefont {Tsirkin}, \citenamefont {Wierzbowska}, \citenamefont
  {Marzari}, \citenamefont {Vanderbilt}, \citenamefont {Souza}, \citenamefont
  {Mostofi},\ and\ \citenamefont {Yates}}]{WANNIER90}%
  \BibitemOpen
  \bibfield  {author} {\bibinfo {author} {\bibfnamefont {G.}~\bibnamefont
  {Pizzi}}, \bibinfo {author} {\bibfnamefont {V.}~\bibnamefont {Vitale}},
  \bibinfo {author} {\bibfnamefont {R.}~\bibnamefont {Arita}}, \bibinfo
  {author} {\bibfnamefont {S.}~\bibnamefont {Blügel}}, \bibinfo {author}
  {\bibfnamefont {F.}~\bibnamefont {Freimuth}}, \bibinfo {author}
  {\bibfnamefont {G.}~\bibnamefont {G{\'{e}}ranton}}, \bibinfo {author}
  {\bibfnamefont {M.}~\bibnamefont {Gibertini}}, \bibinfo {author}
  {\bibfnamefont {D.}~\bibnamefont {Gresch}}, \bibinfo {author} {\bibfnamefont
  {C.}~\bibnamefont {Johnson}}, \bibinfo {author} {\bibfnamefont
  {T.}~\bibnamefont {Koretsune}}, \bibinfo {author} {\bibfnamefont
  {J.}~\bibnamefont {Iba{\~{n}}ez-Azpiroz}}, \bibinfo {author} {\bibfnamefont
  {H.}~\bibnamefont {Lee}}, \bibinfo {author} {\bibfnamefont {J.-M.}\
  \bibnamefont {Lihm}}, \bibinfo {author} {\bibfnamefont {D.}~\bibnamefont
  {Marchand}}, \bibinfo {author} {\bibfnamefont {A.}~\bibnamefont {Marrazzo}},
  \bibinfo {author} {\bibfnamefont {Y.}~\bibnamefont {Mokrousov}}, \bibinfo
  {author} {\bibfnamefont {J.~I.}\ \bibnamefont {Mustafa}}, \bibinfo {author}
  {\bibfnamefont {Y.}~\bibnamefont {Nohara}}, \bibinfo {author} {\bibfnamefont
  {Y.}~\bibnamefont {Nomura}}, \bibinfo {author} {\bibfnamefont
  {L.}~\bibnamefont {Paulatto}}, \bibinfo {author} {\bibfnamefont
  {S.}~\bibnamefont {Ponc{\'{e}}}}, \bibinfo {author} {\bibfnamefont
  {T.}~\bibnamefont {Ponweiser}}, \bibinfo {author} {\bibfnamefont
  {J.}~\bibnamefont {Qiao}}, \bibinfo {author} {\bibfnamefont {F.}~\bibnamefont
  {Thöle}}, \bibinfo {author} {\bibfnamefont {S.~S.}\ \bibnamefont {Tsirkin}},
  \bibinfo {author} {\bibfnamefont {M.}~\bibnamefont {Wierzbowska}}, \bibinfo
  {author} {\bibfnamefont {N.}~\bibnamefont {Marzari}}, \bibinfo {author}
  {\bibfnamefont {D.}~\bibnamefont {Vanderbilt}}, \bibinfo {author}
  {\bibfnamefont {I.}~\bibnamefont {Souza}}, \bibinfo {author} {\bibfnamefont
  {A.~A.}\ \bibnamefont {Mostofi}}, \ and\ \bibinfo {author} {\bibfnamefont
  {J.~R.}\ \bibnamefont {Yates}},\ }\href {\doibase 10.1088/1361-648x/ab51ff}
  {\bibfield  {journal} {\bibinfo  {journal} {Journal of Physics: Condensed
  Matter}\ }\textbf {\bibinfo {volume} {32}},\ \bibinfo {pages} {165902}
  (\bibinfo {year} {2020})}\BibitemShut {NoStop}%
\bibitem [{\citenamefont {Wu}\ \emph {et~al.}(2018)\citenamefont {Wu},
  \citenamefont {Zhang}, \citenamefont {Song}, \citenamefont {Troyer},\ and\
  \citenamefont {Soluyanov}}]{WANNIERTOOLS1}%
  \BibitemOpen
  \bibfield  {author} {\bibinfo {author} {\bibfnamefont {Q.}~\bibnamefont
  {Wu}}, \bibinfo {author} {\bibfnamefont {S.}~\bibnamefont {Zhang}}, \bibinfo
  {author} {\bibfnamefont {H.-F.}\ \bibnamefont {Song}}, \bibinfo {author}
  {\bibfnamefont {M.}~\bibnamefont {Troyer}}, \ and\ \bibinfo {author}
  {\bibfnamefont {A.~A.}\ \bibnamefont {Soluyanov}},\ }\href {\doibase
  https://doi.org/10.1016/j.cpc.2017.09.033} {\bibfield  {journal} {\bibinfo
  {journal} {Computer Physics Communications}\ }\textbf {\bibinfo {volume}
  {224}},\ \bibinfo {pages} {405} (\bibinfo {year} {2018})}\BibitemShut
  {NoStop}%
\bibitem [{\citenamefont {Sancho}\ \emph {et~al.}(1985)\citenamefont {Sancho},
  \citenamefont {Sancho}, \citenamefont {Sancho},\ and\ \citenamefont
  {Rubio}}]{WANNIERTOOLS2}%
  \BibitemOpen
  \bibfield  {author} {\bibinfo {author} {\bibfnamefont {M.~P.~L.}\
  \bibnamefont {Sancho}}, \bibinfo {author} {\bibfnamefont {J.~M.~L.}\
  \bibnamefont {Sancho}}, \bibinfo {author} {\bibfnamefont {J.~M.~L.}\
  \bibnamefont {Sancho}}, \ and\ \bibinfo {author} {\bibfnamefont
  {J.}~\bibnamefont {Rubio}},\ }\href {\doibase 10.1088/0305-4608/15/4/009}
  {\bibfield  {journal} {\bibinfo  {journal} {Journal of Physics F: Metal
  Physics}\ }\textbf {\bibinfo {volume} {15}},\ \bibinfo {pages} {851}
  (\bibinfo {year} {1985})}\BibitemShut {NoStop}%
\bibitem [{\citenamefont {Kosaka}\ \emph {et~al.}(2020)\citenamefont {Kosaka},
  \citenamefont {Michimura}, \citenamefont {Hirabayashi}, \citenamefont
  {Numakura}, \citenamefont {Iizuka}, \citenamefont {Kuwahara},\ and\
  \citenamefont {Uwatoko}}]{EuZn2Ge2}%
  \BibitemOpen
  \bibfield  {author} {\bibinfo {author} {\bibfnamefont {M.}~\bibnamefont
  {Kosaka}}, \bibinfo {author} {\bibfnamefont {S.}~\bibnamefont {Michimura}},
  \bibinfo {author} {\bibfnamefont {H.}~\bibnamefont {Hirabayashi}}, \bibinfo
  {author} {\bibfnamefont {R.}~\bibnamefont {Numakura}}, \bibinfo {author}
  {\bibfnamefont {R.}~\bibnamefont {Iizuka}}, \bibinfo {author} {\bibfnamefont
  {K.}~\bibnamefont {Kuwahara}}, \ and\ \bibinfo {author} {\bibfnamefont
  {Y.}~\bibnamefont {Uwatoko}},\ }\href {\doibase 10.7566/JPSJ.89.054704}
  {\bibfield  {journal} {\bibinfo  {journal} {Journal of the Physical Society
  of Japan}\ }\textbf {\bibinfo {volume} {89}},\ \bibinfo {pages} {054704}
  (\bibinfo {year} {2020})}\BibitemShut {NoStop}%
\bibitem [{\citenamefont {Bud'ko}\ \emph {et~al.}(1999)\citenamefont {Bud'ko},
  \citenamefont {Islam}, \citenamefont {Wiener}, \citenamefont {Fisher},
  \citenamefont {Lacerda},\ and\ \citenamefont {Canfield}}]{RNi2Ge2}%
  \BibitemOpen
  \bibfield  {author} {\bibinfo {author} {\bibfnamefont {S.}~\bibnamefont
  {Bud'ko}}, \bibinfo {author} {\bibfnamefont {Z.}~\bibnamefont {Islam}},
  \bibinfo {author} {\bibfnamefont {T.}~\bibnamefont {Wiener}}, \bibinfo
  {author} {\bibfnamefont {I.}~\bibnamefont {Fisher}}, \bibinfo {author}
  {\bibfnamefont {A.}~\bibnamefont {Lacerda}}, \ and\ \bibinfo {author}
  {\bibfnamefont {P.}~\bibnamefont {Canfield}},\ }\href {\doibase
  https://doi.org/10.1016/S0304-8853(99)00486-2} {\bibfield  {journal}
  {\bibinfo  {journal} {Journal of Magnetism and Magnetic Materials}\ }\textbf
  {\bibinfo {volume} {205}},\ \bibinfo {pages} {53} (\bibinfo {year}
  {1999})}\BibitemShut {NoStop}%
\bibitem [{\citenamefont {{Ramirez}}(1994)}]{Annual1994}%
  \BibitemOpen
  \bibfield  {author} {\bibinfo {author} {\bibfnamefont {A.~P.}\ \bibnamefont
  {{Ramirez}}},\ }\href {\doibase 10.1146/annurev.ms.24.080194.002321}
  {\bibfield  {journal} {\bibinfo  {journal} {Annual Review of Materials
  Research}\ }\textbf {\bibinfo {volume} {24}},\ \bibinfo {pages} {453}
  (\bibinfo {year} {1994})}\BibitemShut {NoStop}%
\bibitem [{\citenamefont {Umeo}\ \emph {et~al.}()\citenamefont {Umeo},
  \citenamefont {Kubo}, \citenamefont {Onimaru}, \citenamefont {Katoh},\ and\
  \citenamefont {Takabatake}}]{JPS}%
  \BibitemOpen
  \bibfield  {author} {\bibinfo {author} {\bibfnamefont {K.}~\bibnamefont
  {Umeo}}, \bibinfo {author} {\bibfnamefont {H.}~\bibnamefont {Kubo}}, \bibinfo
  {author} {\bibfnamefont {T.}~\bibnamefont {Onimaru}}, \bibinfo {author}
  {\bibfnamefont {K.}~\bibnamefont {Katoh}}, \ and\ \bibinfo {author}
  {\bibfnamefont {T.}~\bibnamefont {Takabatake}},\ }in\ \href {\doibase
  10.7566/JPSCP.3.014005} {\emph {\bibinfo {booktitle} {Proceedings of the
  International Conference on Strongly Correlated Electron Systems
  (SCES2013)}}}\BibitemShut {NoStop}%
\bibitem [{\citenamefont {Gibson}\ \emph {et~al.}(1996)\citenamefont {Gibson},
  \citenamefont {Pöttgen}, \citenamefont {Kremer}, \citenamefont {Simon},\
  and\ \citenamefont {Ziebeck}}]{ErAgGe}%
  \BibitemOpen
  \bibfield  {author} {\bibinfo {author} {\bibfnamefont {B.}~\bibnamefont
  {Gibson}}, \bibinfo {author} {\bibfnamefont {R.}~\bibnamefont {Pöttgen}},
  \bibinfo {author} {\bibfnamefont {R.~K.}\ \bibnamefont {Kremer}}, \bibinfo
  {author} {\bibfnamefont {A.}~\bibnamefont {Simon}}, \ and\ \bibinfo {author}
  {\bibfnamefont {K.~R.}\ \bibnamefont {Ziebeck}},\ }\href {\doibase
  https://doi.org/10.1016/0925-8388(96)02201-3} {\bibfield  {journal} {\bibinfo
   {journal} {Journal of Alloys and Compounds}\ }\textbf {\bibinfo {volume}
  {239}},\ \bibinfo {pages} {34} (\bibinfo {year} {1996})}\BibitemShut
  {NoStop}%
\bibitem [{\citenamefont {Talik}\ \emph {et~al.}(2006)\citenamefont {Talik},
  \citenamefont {Kusz}, \citenamefont {Hofmeister}, \citenamefont {Matlak},
  \citenamefont {Skutecka},\ and\ \citenamefont {Klimczak}}]{GdPdX}%
  \BibitemOpen
  \bibfield  {author} {\bibinfo {author} {\bibfnamefont {E.}~\bibnamefont
  {Talik}}, \bibinfo {author} {\bibfnamefont {J.}~\bibnamefont {Kusz}},
  \bibinfo {author} {\bibfnamefont {W.}~\bibnamefont {Hofmeister}}, \bibinfo
  {author} {\bibfnamefont {M.}~\bibnamefont {Matlak}}, \bibinfo {author}
  {\bibfnamefont {M.}~\bibnamefont {Skutecka}}, \ and\ \bibinfo {author}
  {\bibfnamefont {M.}~\bibnamefont {Klimczak}},\ }\href {\doibase
  https://doi.org/10.1016/j.jallcom.2005.12.054} {\bibfield  {journal}
  {\bibinfo  {journal} {Journal of Alloys and Compounds}\ }\textbf {\bibinfo
  {volume} {423}},\ \bibinfo {pages} {47} (\bibinfo {year} {2006})}\BibitemShut
  {NoStop}%
\bibitem [{\citenamefont {Matthias}\ \emph {et~al.}(1963)\citenamefont
  {Matthias}, \citenamefont {Geballe},\ and\ \citenamefont {Compton}}]{Pb}%
  \BibitemOpen
  \bibfield  {author} {\bibinfo {author} {\bibfnamefont {B.~T.}\ \bibnamefont
  {Matthias}}, \bibinfo {author} {\bibfnamefont {T.~H.}\ \bibnamefont
  {Geballe}}, \ and\ \bibinfo {author} {\bibfnamefont {V.~B.}\ \bibnamefont
  {Compton}},\ }\href {\doibase 10.1103/RevModPhys.35.1} {\bibfield  {journal}
  {\bibinfo  {journal} {Rev. Mod. Phys.}\ }\textbf {\bibinfo {volume} {35}},\
  \bibinfo {pages} {1} (\bibinfo {year} {1963})}\BibitemShut {NoStop}%
\bibitem [{\citenamefont {Rossiter}(1987)}]{rossiter_1987}%
  \BibitemOpen
  \bibfield  {author} {\bibinfo {author} {\bibfnamefont {P.~L.}\ \bibnamefont
  {Rossiter}},\ }\href {\doibase 10.1017/CBO9780511600289} {\emph {\bibinfo
  {title} {The Electrical Resistivity of Metals and Alloys}}},\ Cambridge Solid
  State Science Series\ (\bibinfo  {publisher} {Cambridge University Press},\
  \bibinfo {year} {1987})\BibitemShut {NoStop}%
\bibitem [{\citenamefont {Moriya}\ and\ \citenamefont
  {Takimoto}(1995)}]{TohruMoriya1995}%
  \BibitemOpen
  \bibfield  {author} {\bibinfo {author} {\bibfnamefont {T.}~\bibnamefont
  {Moriya}}\ and\ \bibinfo {author} {\bibfnamefont {T.}~\bibnamefont
  {Takimoto}},\ }\href {\doibase 10.1143/jpsj.64.960} {\bibfield  {journal}
  {\bibinfo  {journal} {Journal of the Physical Society of Japan}\ }\textbf
  {\bibinfo {volume} {64}},\ \bibinfo {pages} {960} (\bibinfo {year}
  {1995})}\BibitemShut {NoStop}%
\bibitem [{\citenamefont {Mukhopadhyay}\ \emph {et~al.}(2021)\citenamefont
  {Mukhopadhyay}, \citenamefont {Singh}, \citenamefont {Sen}, \citenamefont
  {Mukherjee}, \citenamefont {Nayak},\ and\ \citenamefont {Mohapatra}}]{RPdSi}%
  \BibitemOpen
  \bibfield  {author} {\bibinfo {author} {\bibfnamefont {A.}~\bibnamefont
  {Mukhopadhyay}}, \bibinfo {author} {\bibfnamefont {K.}~\bibnamefont {Singh}},
  \bibinfo {author} {\bibfnamefont {S.}~\bibnamefont {Sen}}, \bibinfo {author}
  {\bibfnamefont {K.}~\bibnamefont {Mukherjee}}, \bibinfo {author}
  {\bibfnamefont {A.~K.}\ \bibnamefont {Nayak}}, \ and\ \bibinfo {author}
  {\bibfnamefont {N.}~\bibnamefont {Mohapatra}},\ }\href {\doibase
  10.1088/1361-648x/ac1880} {\bibfield  {journal} {\bibinfo  {journal} {Journal
  of Physics: Condensed Matter}\ }\textbf {\bibinfo {volume} {33}},\ \bibinfo
  {pages} {435804} (\bibinfo {year} {2021})}\BibitemShut {NoStop}%
\bibitem [{\citenamefont {Hu}\ \emph {et~al.}(2016)\citenamefont {Hu},
  \citenamefont {Tang}, \citenamefont {Liu}, \citenamefont {Liu}, \citenamefont
  {Zhu}, \citenamefont {Graf}, \citenamefont {Myhro}, \citenamefont {Tran},
  \citenamefont {Lau}, \citenamefont {Wei},\ and\ \citenamefont
  {Mao}}]{ZrSiSe&ZrSiTe}%
  \BibitemOpen
  \bibfield  {author} {\bibinfo {author} {\bibfnamefont {J.}~\bibnamefont
  {Hu}}, \bibinfo {author} {\bibfnamefont {Z.}~\bibnamefont {Tang}}, \bibinfo
  {author} {\bibfnamefont {J.}~\bibnamefont {Liu}}, \bibinfo {author}
  {\bibfnamefont {X.}~\bibnamefont {Liu}}, \bibinfo {author} {\bibfnamefont
  {Y.}~\bibnamefont {Zhu}}, \bibinfo {author} {\bibfnamefont {D.}~\bibnamefont
  {Graf}}, \bibinfo {author} {\bibfnamefont {K.}~\bibnamefont {Myhro}},
  \bibinfo {author} {\bibfnamefont {S.}~\bibnamefont {Tran}}, \bibinfo {author}
  {\bibfnamefont {C.~N.}\ \bibnamefont {Lau}}, \bibinfo {author} {\bibfnamefont
  {J.}~\bibnamefont {Wei}}, \ and\ \bibinfo {author} {\bibfnamefont
  {Z.}~\bibnamefont {Mao}},\ }\href {\doibase 10.1103/PhysRevLett.117.016602}
  {\bibfield  {journal} {\bibinfo  {journal} {Phys. Rev. Lett.}\ }\textbf
  {\bibinfo {volume} {117}},\ \bibinfo {pages} {016602} (\bibinfo {year}
  {2016})}\BibitemShut {NoStop}%
\bibitem [{\citenamefont {Sankar}\ \emph {et~al.}(2017)\citenamefont {Sankar},
  \citenamefont {Peramaiyan}, \citenamefont {Muthuselvam}, \citenamefont
  {Butler}, \citenamefont {Dimitri}, \citenamefont {Neupane}, \citenamefont
  {Rao}, \citenamefont {Lin},\ and\ \citenamefont {Chou}}]{ZrSiS}%
  \BibitemOpen
  \bibfield  {author} {\bibinfo {author} {\bibfnamefont {R.}~\bibnamefont
  {Sankar}}, \bibinfo {author} {\bibfnamefont {G.}~\bibnamefont {Peramaiyan}},
  \bibinfo {author} {\bibfnamefont {I.~P.}\ \bibnamefont {Muthuselvam}},
  \bibinfo {author} {\bibfnamefont {C.~J.}\ \bibnamefont {Butler}}, \bibinfo
  {author} {\bibfnamefont {K.}~\bibnamefont {Dimitri}}, \bibinfo {author}
  {\bibfnamefont {M.}~\bibnamefont {Neupane}}, \bibinfo {author} {\bibfnamefont
  {G.~N.}\ \bibnamefont {Rao}}, \bibinfo {author} {\bibfnamefont {M.-T.}\
  \bibnamefont {Lin}}, \ and\ \bibinfo {author} {\bibfnamefont {F.~C.}\
  \bibnamefont {Chou}},\ }\href {\doibase 10.1038/srep40603} {\bibfield
  {journal} {\bibinfo  {journal} {Scientific Reports}\ }\textbf {\bibinfo
  {volume} {7}},\ \bibinfo {pages} {40603} (\bibinfo {year}
  {2017})}\BibitemShut {NoStop}%
\bibitem [{\citenamefont {Laha}\ \emph {et~al.}(2019)\citenamefont {Laha},
  \citenamefont {Malick}, \citenamefont {Singha}, \citenamefont {Mandal},
  \citenamefont {Rambabu}, \citenamefont {Kanchana},\ and\ \citenamefont
  {Hossain}}]{YbCdGe}%
  \BibitemOpen
  \bibfield  {author} {\bibinfo {author} {\bibfnamefont {A.}~\bibnamefont
  {Laha}}, \bibinfo {author} {\bibfnamefont {S.}~\bibnamefont {Malick}},
  \bibinfo {author} {\bibfnamefont {R.}~\bibnamefont {Singha}}, \bibinfo
  {author} {\bibfnamefont {P.}~\bibnamefont {Mandal}}, \bibinfo {author}
  {\bibfnamefont {P.}~\bibnamefont {Rambabu}}, \bibinfo {author} {\bibfnamefont
  {V.}~\bibnamefont {Kanchana}}, \ and\ \bibinfo {author} {\bibfnamefont
  {Z.}~\bibnamefont {Hossain}},\ }\href {\doibase 10.1103/PhysRevB.99.241102}
  {\bibfield  {journal} {\bibinfo  {journal} {Phys. Rev. B}\ }\textbf {\bibinfo
  {volume} {99}},\ \bibinfo {pages} {241102} (\bibinfo {year}
  {2019})}\BibitemShut {NoStop}%
\bibitem [{\citenamefont {Roy}\ \emph {et~al.}(2021)\citenamefont {Roy},
  \citenamefont {Singha}, \citenamefont {Ghosh},\ and\ \citenamefont
  {Mandal}}]{Ta3SiTe6}%
  \BibitemOpen
  \bibfield  {author} {\bibinfo {author} {\bibfnamefont {S.}~\bibnamefont
  {Roy}}, \bibinfo {author} {\bibfnamefont {R.}~\bibnamefont {Singha}},
  \bibinfo {author} {\bibfnamefont {A.}~\bibnamefont {Ghosh}}, \ and\ \bibinfo
  {author} {\bibfnamefont {P.}~\bibnamefont {Mandal}},\ }\href {\doibase
  10.1103/PhysRevMaterials.5.064203} {\bibfield  {journal} {\bibinfo  {journal}
  {Phys. Rev. Materials}\ }\textbf {\bibinfo {volume} {5}},\ \bibinfo {pages}
  {064203} (\bibinfo {year} {2021})}\BibitemShut {NoStop}%
\bibitem [{\citenamefont {Hyart}\ \emph {et~al.}(2018)\citenamefont {Hyart},
  \citenamefont {Ojaj{\"a}rvi},\ and\ \citenamefont {Heikkil{\"a}}}]{Berry}%
  \BibitemOpen
  \bibfield  {author} {\bibinfo {author} {\bibfnamefont {T.}~\bibnamefont
  {Hyart}}, \bibinfo {author} {\bibfnamefont {R.}~\bibnamefont {Ojaj{\"a}rvi}},
  \ and\ \bibinfo {author} {\bibfnamefont {T.~T.}\ \bibnamefont
  {Heikkil{\"a}}},\ }\href {\doibase 10.1007/s10909-017-1846-3} {\bibfield
  {journal} {\bibinfo  {journal} {Journal of Low Temperature Physics}\ }\textbf
  {\bibinfo {volume} {191}},\ \bibinfo {pages} {35} (\bibinfo {year}
  {2018})}\BibitemShut {NoStop}%
\bibitem [{\citenamefont {Laha}\ \emph {et~al.}(2020)\citenamefont {Laha},
  \citenamefont {Rambabu}, \citenamefont {Kanchana}, \citenamefont {Petit},
  \citenamefont {Szotek},\ and\ \citenamefont {Hossain}}]{YbCdSn}%
  \BibitemOpen
  \bibfield  {author} {\bibinfo {author} {\bibfnamefont {A.}~\bibnamefont
  {Laha}}, \bibinfo {author} {\bibfnamefont {P.}~\bibnamefont {Rambabu}},
  \bibinfo {author} {\bibfnamefont {V.}~\bibnamefont {Kanchana}}, \bibinfo
  {author} {\bibfnamefont {L.}~\bibnamefont {Petit}}, \bibinfo {author}
  {\bibfnamefont {Z.}~\bibnamefont {Szotek}}, \ and\ \bibinfo {author}
  {\bibfnamefont {Z.}~\bibnamefont {Hossain}},\ }\href {\doibase
  10.1103/PhysRevB.102.235135} {\bibfield  {journal} {\bibinfo  {journal}
  {Phys. Rev. B}\ }\textbf {\bibinfo {volume} {102}},\ \bibinfo {pages}
  {235135} (\bibinfo {year} {2020})}\BibitemShut {NoStop}%
\bibitem [{\citenamefont {Emmanouilidou}\ \emph {et~al.}(2017)\citenamefont
  {Emmanouilidou}, \citenamefont {Shen}, \citenamefont {Deng}, \citenamefont
  {Chang}, \citenamefont {Shi}, \citenamefont {Kotliar}, \citenamefont {Xu},\
  and\ \citenamefont {Ni}}]{CaTX}%
  \BibitemOpen
  \bibfield  {author} {\bibinfo {author} {\bibfnamefont {E.}~\bibnamefont
  {Emmanouilidou}}, \bibinfo {author} {\bibfnamefont {B.}~\bibnamefont {Shen}},
  \bibinfo {author} {\bibfnamefont {X.}~\bibnamefont {Deng}}, \bibinfo {author}
  {\bibfnamefont {T.-R.}\ \bibnamefont {Chang}}, \bibinfo {author}
  {\bibfnamefont {A.}~\bibnamefont {Shi}}, \bibinfo {author} {\bibfnamefont
  {G.}~\bibnamefont {Kotliar}}, \bibinfo {author} {\bibfnamefont {S.-Y.}\
  \bibnamefont {Xu}}, \ and\ \bibinfo {author} {\bibfnamefont {N.}~\bibnamefont
  {Ni}},\ }\href {\doibase 10.1103/PhysRevB.95.245113} {\bibfield  {journal}
  {\bibinfo  {journal} {Phys. Rev. B}\ }\textbf {\bibinfo {volume} {95}},\
  \bibinfo {pages} {245113} (\bibinfo {year} {2017})}\BibitemShut {NoStop}%
\end{thebibliography}%

\end{document}